\documentclass[paper]{JHEP3} 

\JHEPspecialurl{http://jhep.sissa.it/JOURNAL/JHEP3.tar.gz}

\usepackage{epsfig,multicol}

\newcommand\fverb{\setbox\pippobox=\hbox\bgroup\verb}
\newcommand\fverbdo{\egroup\medskip\noindent%
			\fbox{\unhbox\pippobox}\ }
\newcommand\fverbit{\egroup\item[\fbox{\unhbox\pippobox}]}
\newcommand{\id}{{1\!\!1}} 
\newcommand {\beq}{\begin{equation}}
\newcommand {\eeq}{\end{equation}}
\newcommand {\beqa}{\begin{eqnarray}}
\newcommand {\eeqa}{\end{eqnarray}}
\newcommand {\n}{\nonumber \\}
\newcommand {\tr}{{\rm tr\,}}

\newbox\pippobox

\title{Testing the Gaussian expansion method\\
in exactly solvable matrix models}

\author{Jun Nishimura\\
High Energy Accelerator Research Organization (KEK)\\
1-1 Oho, Tsukuba 305-0801, Japan\\
E-mail: \email{jnishi@post.kek.jp}}
\author{Toshiyuki Okubo\\
Department of Physics, Nagoya University\\
Furo-cho, Chikusa-ku, Nagoya 464-8602, Japan\\
E-mail: \email{okubo@eken.phys.nagoya-u.ac.jp}}
\author{Fumihiko Sugino\\
School of Physics \& BK-21 Physics Division\\
Seoul National University, 
Seoul 151-747, Korea\\
E-mail: \email{sugino@phya.snu.ac.kr}}

\preprint{KEK-TH-918\\DPNU-03-23\\SNUST-030802\\\hepth{0309262}}	

\abstract{
The Gaussian expansion has been developed since early 80s as
a powerful analytical method, which enables {\em nonperturbative} studies
of various systems using `perturbative' calculations.
Recently the method has been used to suggest that
4d space-time is generated dynamically
in a matrix model formulation of superstring theory.
Here we clarify the nature of the method by applying it to exactly
solvable one-matrix models with various kinds of potential including 
the ones unbounded from below and of the double-well type.
We also formulate a prescription to include a linear term in the Gaussian
action in a way consistent with the loop expansion,
and test it in some concrete examples.
We discuss a case where we obtain two distinct plateaus in the parameter
space of the Gaussian action, corresponding to different large-$N$
solutions. This clarifies the situation encountered
in the dynamical determination 
of the space-time dimensionality in the previous works.
}


\begin{document} 


\section{Introduction}

The Gaussian expansion is a powerful nonperturbative
method, which has been applied to condensed matter physics
and statistical physics extensively.
Even at the lowest order, which is sometimes referred to as
the self-consistent harmonic approximation,
it allows us to understand fundamental properties of various 
systems at least qualitatively.
In fact there exists a systematic way to improve the
approximation, which we refer to as the Gaussian expansion method,
but it is also called in the literature under various names
such as improved mean field approximation, delta expansion and so on.
The method has been worked out first in
quantum mechanical systems \cite{conv,Stevenson:1981vj},
where the expansion was shown to be convergent 
in some concrete examples \cite{exact_conv},
and it has been generalized to field theory later \cite{GEM_field}.
(The basic idea is also used in optimizing perturbation theory
where the results depend on the renormalization scheme 
\cite{Stevenson:1981vj,Dhar:sh,Kawamoto:2003kn}.)
The most peculiar feature of this method, as emphasized by 
Stevenson \cite{Stevenson:1981vj},
is that one obtains genuinely
nonperturbative results --- since never in the whole procedure
does one attempt an expansion with respect to the coupling
constant --- and yet the required task is nothing more than
familiar perturbative calculations based on Feynman diagrams.

Recently this method has been applied to superstring/M theories
using their matrix model formulations.
In Ref.\ \cite{Kabat:2000hp} Kabat and Lifschytz proposed
to use the Gaussian approximation
in the Matrix Theory \cite{Banks:1996vh},
which is conjectured to be a nonperturbative definition of M-theory
in the infinite momentum frame.
Indeed Refs.\ \cite{blackholes} were able to
reveal interesting blackhole thermodynamics from the dual strong-coupling
gauge theory.
An earlier application of the Gaussian approximation to 
random matrix models can be found in Ref.\ \cite{EL}.

In Ref.\ \cite{Nishimura:2001sx} two of the present authors (J.N.\ and F.S.) 
applied the Gaussian expansion method to the IIB matrix model \cite{IKKT},
which is conjectured to be a nonperturbative definition of type IIB
superstring theory in 10 dimensions.
One of the most interesting questions in this model
\footnote{
The finiteness of the partition function has been
proved in Refs.\ \cite{Austing:2001bd}.
Various simplified versions were studied
by Monte Carlo simulations \cite{HNT,monte}
and by the Gaussian expansion method \cite{Gauss_simpleIIB}.
}
concerns the possibility that the 4d space-time \cite{Aoki:1998vn} 
appears {\em dynamically} accompanied with the SSB of the SO(10) symmetry 
down to SO(4). 
From the path-integral point of view,
this phenomenon may be caused
by the phase of the fermion determinant \cite{NV},
and Monte Carlo results support this mechanism \cite{sign}
(Other possible mechanisms are discussed in 
Refs.\ \cite{Aoki:1998vn,Vernizzi:2002mu}).
In Ref.\ \cite{Nishimura:2001sx} the same
issue has been addressed analytically by using the Gaussian expansion method.
The main idea was to consider various kinds of Gaussian action preserving 
only some subgroup of SO(10) 
and to identify the `true vacuum' by comparing the corresponding free energy.
Calculations up to the 3rd order showed that
space-time preserving the SO(4) symmetry
has the smallest free energy, and that
the ratio of the extents in the four directions and the remaining
six directions increases with the order.
To our knowledge, this is the first time in history
that the space-time dimensionality `4' is suggested
from the nonperturbative dynamics of superstring theory.
\footnote{Recently another indication of this phenomenon has been
obtained from the calculations of the 2-loop effective action 
around fuzzy-sphere solutions \cite{Imai:2003jb}.}


A lot of effort has then been made to increase the order of the expansion.
In Ref.\ \cite{KKKMS} it was noticed that 
Schwinger-Dyson equations can be used to reduce considerably
the number of Feynman diagrams to be evaluated.
In Ref.\ \cite{Kawai:2002ub} a computer code 
has been written in order to automatize the task of
listing up and evaluating all the Feynman diagrams.
With these technical developments, the order of the expansion
has now been increased up to the 7th order, and the results
strengthened the conclusion of Ref.\ \cite{Nishimura:2001sx}.
The nature of the method itself has also been clarified.
In Ref.\ \cite{KKKMS} the Gaussian expansion method was interpreted
as an improved Taylor expansion,
and the importance of identifying a plateau in the space of
free parameters in the Gaussian action was recognized.

Although these new results in matrix models obtained
by the Gaussian expansion method are quite encouraging and deserve
further investigations, 
we consider it equally important to know the method itself
by applying it to a well-understood system and to
try to improve or refine the method.
In our previous work \cite{Nishimura:2002va} we applied the method to
the bosonic version of the IIB matrix model,
where Monte Carlo results \cite{HNT} are available,
and the convergence of the method has been demonstrated.
We also developed a new technique to deal with the free parameters
in the Gaussian action, which were conventionally
determined by solving the `self-consistency equations'.

In this paper we attempt to gain more experience with the method
by applying it to exactly solvable one-matrix models 
\cite{Brezin:1977sv},
which have been studied intensively in the context of two-dimensional
quantum gravity and non-critical string theory.
The Gaussian expansion has been carried out maximally up to 
the 18th order
and the results are compared with the known exact results.

First we study the $\phi^4$ matrix model 
with various kinds of potential including 
the ones unbounded from below \cite{Brezin:1977sv}
and of the double-well type 
\cite{Cicuta:1986pu}.
For the unbounded potential, it is known that a stable vacuum exists
in the large-$N$ limit if the parameters of the potential
satisfy a certain condition.
In this case the Gaussian expansion converges 
and reproduces the exact results accurately. 
If the parameters are chosen such that there is no stable vacuum even
in the large-$N$ limit,
the Gaussian expansion does not converge either.
Thus the method captures correctly
the critical phenomenon associated with
the overflow of the eigenvalues.
For the double-well potential, on the other hand, we find that 
the results are in reasonable agreement with the exact values
at low orders, but they start to oscillate violently as we go to higher 
orders. 

Next we consider the $\phi^3$ matrix model.
In this case the potential is always unbounded from below,
but again a stable vacuum is known to exist
in the large-$N$ limit if the parameters of the potential are chosen
appropriately.
A new ingredient here is that 
the model does not have the Z$_2$ symmetry unlike 
the $\phi^4$ matrix model, and therefore
we have to include a linear term in the Gaussian action.
We formulate a prescription to deal with the linear term
in a way consistent with the loop expansion.
Since we have two free parameters 
in the Gaussian action in this case,
identifying a plateau in the parameter space becomes more
nontrivial than in the $\phi^4$ matrix model.
Here we find the histogram technique proposed in 
our previous paper \cite{Nishimura:2002va}
to be quite useful, 
and we do obtain results converging to the exact values.

Finally we reconsider the $\phi^4$ matrix model with the double-well 
potential, in which the Gaussian expansion method in its simplest
form seems to fail as mentioned above.
Here we redo the analysis with the inclusion of a linear term 
in the Gaussian action. Two types of linear term are considered,
and for each case we obtain a plateau with different free energy.
When we use a SU($N$) invariant but Z$_2$ breaking linear term, 
we obtain a plateau which corresponds to putting all the eigenvalues 
into a single well.
When we use a linear term which has the `Z$_2$ symmetry'
but breaks the SU($N$) symmetry down to SU($N/2$)$\times$SU($N/2$),
we obtain a plateau which corresponds to partitioning
the eigenvalues equally into the two wells.
The free energy for the latter is smaller 
in accord with exact results.
This case therefore provides an example where we have more than one plateaus
corresponding to different large-$N$ saddle-point solutions,
but the true vacuum can still be determined by identifying the plateau 
which gives the smallest free energy. This supports the strategy taken
to determine the dynamical space-time dimensionality
in the aforementioned works in the IIB matrix model
\cite{Nishimura:2001sx,KKKMS,Kawai:2002ub}. 

This paper is organized as follows.
In Section \ref{section:phi4} we study the 
$\phi^4$ matrix model with various types of potential.
In Section \ref{section:linear} we 
study the $\phi^3$ matrix model, where
we discuss how to treat a linear term in the Gaussian action
systematically.
In Section \ref{section:double-well-revisited}
we revisit the $\phi^4$ matrix model with the double-well potential
including a linear term in the Gaussian action.
Section \ref{section:summary} is devoted to summary and discussions.
In Appendix \ref{sec:asym2cut} we discuss general large-$N$ 
saddle-point solutions in the $\phi^4$ matrix model with the double-well 
potential.
In Appendices \ref{sec:mixed} and \ref{sec:SDmethod}
we present the details of the calculations we have done in
Section \ref{section:double-well-revisited}.


\section{$\phi^4$ matrix model}
\label{section:phi4}

In this Section we apply the Gaussian expansion method
to the $\phi^4$ matrix model,
which is exactly solvable and is known to exhibit 
nontrivial critical phenomena.
We also give a brief review of the method
in Section \ref{basic_idea}.

\subsection{Exact solutions}
\label{exact_solutions}

The $\phi^4$ matrix model we study in this Section
is defined by the partition function
\begin{eqnarray}
  Z &=& \int d^{N^2}\phi \, e^{-S}, \label{phi4Z} \\
  S &=& N\frac{m}{2} \tr \phi^2 + N\frac{g}{4} \tr\phi^4, 
\label{phi4S}
\end{eqnarray}
where $\phi$ is a $N\times N$ hermitian matrix. 
The integration measure $d^{N^2}\phi$
is defined as $d^{N^2}\phi = \prod_{a=1}^{N^2} d\phi^a/{\sqrt{2\pi}}$, 
where $\phi^a$
are the coefficients in the expansion $\phi = \sum_a \phi^a T^a$ 
with respect to the U($N$) generators $T^a$ normalized as
$\tr (T^a T^b) = (1/2)\delta^{ab}$.
By rescaling the variables $\phi$ as $\phi \mapsto \sqrt{|m|}^{-1}\phi$,
one finds that
the system can be characterized by the sign of $m$ and the 
effective coupling constant $g_{\rm eff}=g/m^2$.

In order to discuss the property of this model,
it is convenient to look at the eigenvalues of the matrix $\phi$ .
Let us therefore diagonalize $\phi$ as
\beq
\phi = V \Lambda V^{\dag} \ ,
\eeq
where $V$ is a unitary matrix and 
$\Lambda = {\rm diag}(\lambda_1 , \cdots , \lambda_N)$ 
is a real diagonal matrix. 
Integrating out the `angular variables' $V$,
one obtains the effective action for 
the eigenvalues $\lambda_i$ as
\beqa
\label{eff_action}
S_{\rm eff} &=& \sum_{i=1}^{N} V(\lambda_i)
- \sum_{i\neq j} \ln |\lambda_i - \lambda_j|  \ , \\
V(x) &=& N\left( \frac{m}{2} x^2 + \frac{g}{4} x^4 \right)  \ ,
\label{potential}
\eeqa
where the second term in (\ref{eff_action})
comes from the measure $d^{N^2}\phi$.
Thus the effective theory for the eigenvalues
may be regarded as a system of $N$ particles
in the potential (\ref{potential})
with a logarithmic repulsive force between each pair.
As a fundamental object, we introduce the eigenvalue distribution
\beq
\rho(x) = 
 \left\langle 
\frac1N \sum_{i=1}^{N}
\delta(x-\lambda_i)
\right\rangle  \ .
\label{defrho}
\eeq
Let us also define the free energy $F = - \ln Z$
and the observable $\left\langle \frac1N \tr \phi^\ell \right\rangle$,
which can be written in terms of $\rho(x)$ as
\beq
\left\langle \frac1N \tr \phi^\ell \right\rangle 
= \int \, d x \, x^\ell \rho(x)  \ .
\eeq
\FIGURE{\epsfig{file=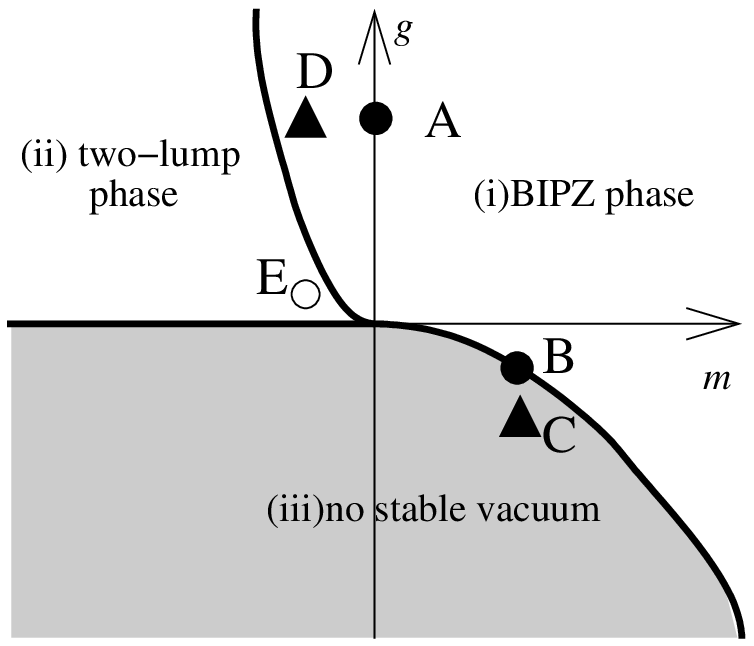,height=8cm} 
    \caption{The phase diagram of the $\phi^4$ matrix model.
The 5 points labeled as A, B, $\cdots$, E denote the set of parameters, 
for which we tested the Gaussian expansion method. 
}
    \label{fig:phase_diagram}
}

In the large-$N$ limit the phase diagram of the model consists
of three regions : 
(i) the BIPZ phase, (ii) the two-lump phase, and (iii) no stable vacuum,
as depicted in Fig.\ \ref{fig:phase_diagram}.

Let us first consider the case $m>0$ and $g>0$,
in which the eigenvalue distribution $\rho(x)$ is given 
by \cite{Brezin:1977sv}
\beq
\rho(x) = 
\frac{1}{\pi}
\left( \frac{1}{2}g x^2 + g a^2 + \frac{1}{2}m
\right)\sqrt{4 a^2 - x^2} 
\label{eigen_BIPZ}
\eeq
for $|x|\le 2a$, where $a$ is defined by
\beq
 a^2 = \frac{2}{m + \sqrt{m^2 + 12g}} \ ,
\label{a2def}
\eeq
and $\rho(x)=0$ otherwise.
Note that $\rho(x)$ has a finite compact support $[-2a, 2a]$
(We remind the reader that the large-$N$ limit is already taken here).
The absence of penetration into the region $|x|>2a$
is due to the fact that the potential (\ref{potential}) grows
linearly as $N$ goes to infinity,
while the distribution does not collapse to the potential minimum
due to the repulsive force between the $N$ particles.
The free energy and the observables
are given by
\begin{eqnarray}
  F(m,g) &=& N^2\left[\frac12 \ln N - \frac{1}{24}(ma^2 - 1)(ma^2 - 9)
    -\frac12 \ln (2a^2) 
+ O\left(\frac{\ln N}{N}\right) \right], \\
\left\langle \frac1N \tr \phi^2 \right\rangle &=& 
\frac13 a^2 (4 -ma^2 ) \mbox{~~~},\mbox{~~~}
 \left\langle \frac1N \tr \phi^4 \right\rangle = 
\frac{a^4}{g}(3 - m a^2 ) \ .
\label{BIPZphase1-phi4} 
\end{eqnarray}

Let us set $m$ to some positive value, and see what happens 
if one decreases $g$.
Note that nothing singular happens at $g=0$,
and the above expressions remain valid for $g<0$ as long
as $m^2 + 12g$ inside the square root in eq.\ (\ref{a2def}) is 
positive. Although the action becomes unbounded,
the potential barrier grows with $N$ fast enough to prevent
the eigenvalues from overflowing.
For $g < - m^2/12$, however, the stable vacuum ceases to exist
due to the overflow of the eigenvalues.
At the critical point $g = - m^2/12$, 
the non-analyticity of $\rho(x)$ at the 
edge of the support ($x_0=\pm 2a$) changes from $|x-x_0|^{1/2}$
to $|x-x_0|^{3/2}$.
This critical phenomenon plays a crucial role 
in the context of two-dimensional quantum gravity and 
non-critical string theory.

Let us next set $g$ to some positive value
and see what happens if one decreases $m$.
From (\ref{eigen_BIPZ}) one finds that $\rho '' (0)$
changes from negative to positive at $m = 2 \sqrt{g}$ ,
meaning that $\rho(x)$ starts to develop a double-peak structure.
This occurs before $m$ reaches zero
due to the repulsive force between the eigenvalues.
From (\ref{eigen_BIPZ}) one also finds that $\rho(0)$ decreases as
one decreases $m$ and 
it finally becomes zero at $m = - 2 \sqrt{g}$.
Beyond this point the eigenvalue distribution $\rho(x)$
will have two compact supports, and it is given explicitly 
by \cite{Cicuta:1986pu}
\beq
\rho(x) = 
\frac{1}{2 \pi} g |x|
\sqrt{(x^2-A_{-}^2)(A_{+}^2 - x^2)} 
\label{splitphase2}
\eeq
for $A_{-} \le |x| \le A_{+}$,
where $A_{\pm}$ is given by
$ A_{\pm}^2 = (|m| \pm 2\sqrt{g})/g$
and $\rho(x) = 0$ otherwise.
At $m = - 2 \sqrt{g}$ 
the distribution (\ref{splitphase2}) reduces to (\ref{eigen_BIPZ}).
The free energy and the observables are given by \cite{Cicuta:1986pu}
\beqa
  F(m,g) &=& N^2\left[\frac12 \ln N - \frac{3}{8}
- \frac{1}{2} \ln 2 - \frac{1}{4} \frac{m^2}{g}
+ \frac{1}{4} \ln g + O\left(\frac{\ln N}{N}\right) \right]  \ ,
\label{splitphase1} 
\\
 \left\langle \frac1N \tr \phi^2 \right\rangle &=& 
\frac{|m|}{g}  \mbox{~~~},\mbox{~~~}
 \left\langle \frac1N \tr \phi^4 \right\rangle = 
\frac{m^2}{g^2} +\frac{1}{g} \ .
\label{splitphase1-phi4} 
\eeqa

We call the phase that continues from the region $m>0$ and $g>0$
as the BIPZ phase, and the phase characterized by the two compact
supports of the eigenvalue distribution as the two-lump phase.


\subsection{The Gaussian expansion method}

As we have seen in the previous Section, the simple one-matrix model
exhibits nontrivial critical phenomena associated with the
eigenvalue distribution. These phenomena are certainly nonperturbative
in the sense that it cannot be seen by perturbative expansion
with respect to $g_{\rm eff}=g/m^2$.
The aim of this Section is to apply the Gaussian expansion method
to this model and to see if it reproduces 
these critical phenomena correctly.

\subsubsection{A brief review}
\label{basic_idea}

We will first illustrate the idea of the Gaussian expansion method
using the $\phi^4$ matrix model as an example.
Let us denote the quadratic term and the quartic term
in the action (\ref{phi4S}) as
$S_2$ and $S_4$ so that $S = S_2 + S_4$.
Ordinary loop expansion can be formulated as follows :
(i) consider the action 
\beq
S_{\rm PT}(\lambda) = \frac{1}{\lambda}(S_2 + S_4) \ ,
\label{pert_action}
\eeq
(ii) calculate free energy and observables 
as an expansion with respect to $\lambda$ up to some finite order,
and (iii) set $\lambda = 1$.
Equivalence to
the perturbative expansion with respect to $g_{\rm eff}=g/m^2$
can be easily seen by rescaling the variables as $\phi \mapsto 
\sqrt{\lambda/|m|}\, \phi$.
The Gaussian expansion method simply amounts to replacing
the action (\ref{pert_action}) by
\beq
S_{\rm GEM} (t ; \lambda)
= \, \frac{1}{\lambda} \, \Bigl[ \Bigl\{ S_0 + \lambda (S_2 - S_0) 
\Bigr\}  + S_4  \Bigr] \ ,
\label{GEM_action}
\eeq
where $S_0$ is taken to be the SU($N$) invariant Gaussian action
\footnote{A Gaussian
action which breaks the SU($N$) symmetry
is considered in Section \ref{vacuum-split}.}
\begin{equation}
  S_0 = \frac{Nt}{2}\tr \phi^2 
\label{gaussian_action}
\end{equation}
with a real positive parameter $t$, which is 
left arbitrary at this point.
Note that both (\ref{pert_action}) and (\ref{GEM_action})
reduces to the original action (\ref{phi4S})
if one sets $\lambda = 1$.
The Gaussian expansion method can also be viewed as a loop expansion but
with the `classical action' $(S_0 + S_4)$ 
and the `one-loop counterterms' $(S_2 - S_0)$.
Note, however, that the method {\em cannot} be viewed as an expansion 
with respect to any parameter in the original action
unlike (\ref{pert_action}),
except when one sets $t=m$, for which one simply retrieves 
the ordinary perturbation theory.
In fact the freedom in changing the parameter $t$ in
the Gaussian action (\ref{gaussian_action})
plays a crucial role in this method.

 
To make things more transparent,
let us rescale $\phi$ as
$\phi \mapsto \sqrt{\lambda}\, \phi$
so that the partition function takes the form
\beqa
Z &=& \int d^{N^2}\phi
\, e^{ - ( S_{\rm cl} + S_{\rm c.t.}) } \ ,  \\
S_{\rm cl} (t;\lambda) &=& S_0 + \lambda S_4   ~~~,~~~
S_{\rm c.t.} (t;\lambda) =  \lambda (S_2 - S_0 )  \ .
\eeqa
Thus in the present example the Gaussian expansion can
be viewed also as an expansion with respect to $(S-S_0)$, 
the difference of the original action
from the Gaussian action.
However this is peculiar to the models with actions containing only
quadratic and quartic terms, and it is not the case in general 
(See Section \ref{section:linear}).
In earlier formulation \cite{Kabat:2000hp,Nishimura:2001sx}
the expansion was first made with respect to $(S-S_0)$,
and then it was reorganized 
in such a way that it becomes consistent with the loop expansion.
Here we have chosen to take a shorter path.

In actual calculations, 
the `one-loop counter terms' can be 
incorporated easily by noticing the relation
\beq
S_{\rm cl} (t;\lambda)  + 
S_{\rm c.t.} (t;\lambda) 
= S_{\rm cl} (t + \lambda(m-t) ; \lambda ) \ .
\eeq
Namely the calculation of a physical quantity
in the Gaussian expansion proceeds in two steps;
(i) obtain the $\lambda$-expansion of the quantity
using the `classical action' $S_{\rm cl} (t ;\lambda)$,
(ii) shift the parameter $t$ by $\lambda(m-t)$
and reorganize the expansion with respect to $\lambda$.
In general the first step can be done by ordinary Feynman diagrammatic
calculations, where the use of Schwinger-Dyson equations
reduces the number of diagrams considerably \cite{KKKMS}.
The large-$N$ limit can be taken, if one wishes,
by simply retaining the planar diagrams only. 
(This results in considerable simplification in the matrix model
applications, but it is definitely not the main point 
of the Gaussian expansion method itself.)
In fact in the present case where we know the
exact results for the `classical action' $S_{\rm cl} (t ;\lambda)$ 
\cite{Brezin:1977sv},
the first step can be done with little effort using softwares 
for symbolic manipulations like Mathematica.
This allows us to test the Gaussian expansion method at very high orders
(we went maximally up to the 18th order).


Let us move on to the results obtained by the Gaussian expansion method.
We compute the `free energy density' defined by
\begin{equation}
  f = \lim_{N\to\infty} \left\{ \frac{F}{N^2} - \frac{1}{2}\ln N\right\} \ ,
\label{fe_density}
\end{equation}
where the second term is subtracted in order to make the quantity finite
in the large-$N$ limit.
For illustrative purposes, 
we consider the case $g=1$, $m=0$ 
(the point A in Fig.\ \ref{fig:phase_diagram}),
which was also studied in Refs.\ \cite{KKKMS,Kawai:2002ub}.
Let us recall that the expansion parameter in ordinary perturbation
theory is $g_{\rm eff} = g/m^2$ 
and the convergence radius is known to be $1/12$.
Therefore the above choice of parameters
corresponds to the strong coupling limit,
which is of course beyond the reach
of ordinary perturbation theory.

\FIGURE{\epsfig{file=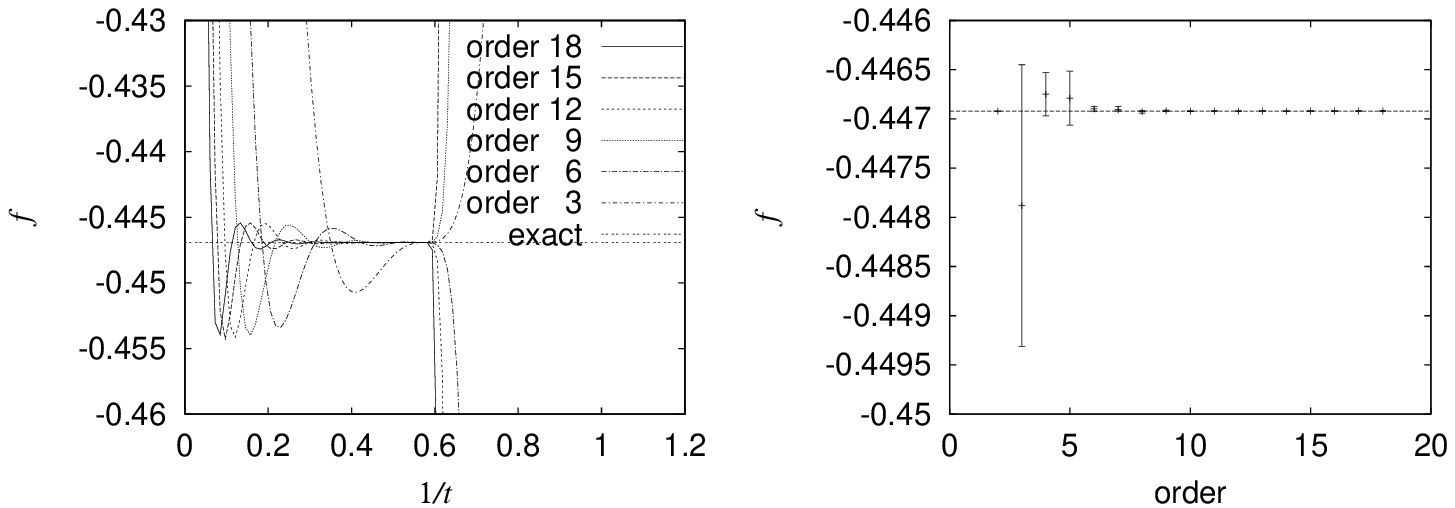,height=5cm} 
    \caption{
{\bf (Left)} The free energy density $f$
calculated by the Gaussian expansion method 
is plotted as a function of $1/t$ for $g = 1$, $m = 0$.
Each curve corresponds to the order 3, 6, 9, 12, 15, 18.
The horizontal line represents the exact value
$f = -3/8 - \log(2/\sqrt{3})/2 = -0.44692 \cdots$.
{\bf (Right)} The free energy density $f$ 
extracted at the order 2 , $\cdots$, 18
by the histogram technique from the results of the 
Gaussian expansion for various $1/t$.
(When we make a histogram, we used the region $0.01 \le  1/t \le  1$.)
The `error bars' represent the theoretical uncertainty estimated
by the same technique.
The order 1 result ($f=-0.423287$) lies outside the displayed region.
The horizontal line represents the exact result.
}
    \label{fig:fe_np_error}
}

In Fig.\ \ref{fig:fe_np_error} (left) we plot the
free energy obtained at the order 3, 6, 9, 12, 15, 18
as a function of $1/t$.
As one can see also from the figure, 
the result of the Gaussian expansion
generically depends on the free parameter $t$
in the Gaussian action (\ref{gaussian_action}).
However, since $t$ is a parameter which is introduced by hand, 
the result should not depend much on it in the region of $t$
where the expansion becomes valid, if such a region exists at all.
Indeed at sufficiently high orders we observe the formation of a plateau,
meaning that there exists a certain range of $t$ 
where the result becomes almost independent of $t$.
Moreover the height of the plateau agrees very accurately with the
exact result represented by the horizontal line.
In old literature the free parameter ($t$ in the present case)
was determined in such a way that the result becomes most insensitive
to the change of the parameter. This strategy was named as 
`the principle of minimum sensitivity' \cite{Stevenson:1981vj}.
The importance of the formation of a plateau has been first recognized
in Ref.\ 
\cite{KKKMS}.



Given this new insight, we still have to develop
some technique to identify a plateau and to extract its height
in order to make a concrete prediction.
Note also that the plateau is not completely flat but has
some small fluctuations in practice.
This causes some theoretical uncertainty in the prediction from
the method, and it is desirable to be able to estimate its order
of magnitude.
In Ref.\ \cite{Nishimura:2002va}
we have developed a technique based on histograms.
Fig.\ \ref{fig:fe_np_error} (right) shows the results obtained 
for the present case.
The `error bars' represent an estimate for the theoretical 
uncertainty explained above. 
Clearly the result converges to the exact value 
within the first several orders.


\subsubsection{Unbounded potential ($g<0$, $m>0$)}
\label{unbounded}

Let us move on to the $g<0$, $m>0$ case, for which the potential 
becomes unbounded from below.
As we discussed in Section \ref{exact_solutions},
a stable vacuum continues to exist for $g\ge -m^2/12$
in the large-$N$ limit.
In Fig.\ \ref{fig:fe_unbound} (left) 
we show the free energy
as a function of $1/t$ 
at the critical point $g = -1/12$, $m = 1$ 
(the point B in Fig.\ \ref{fig:phase_diagram}).
We observe the formation of a clear plateau.
The results for $g > -m^2/12$ were equally successful.
It is noteworthy that the Gaussian expansion method
can reproduce the large-$N$ solution even in the case of 
the unbounded potential.

\FIGURE{\epsfig{file=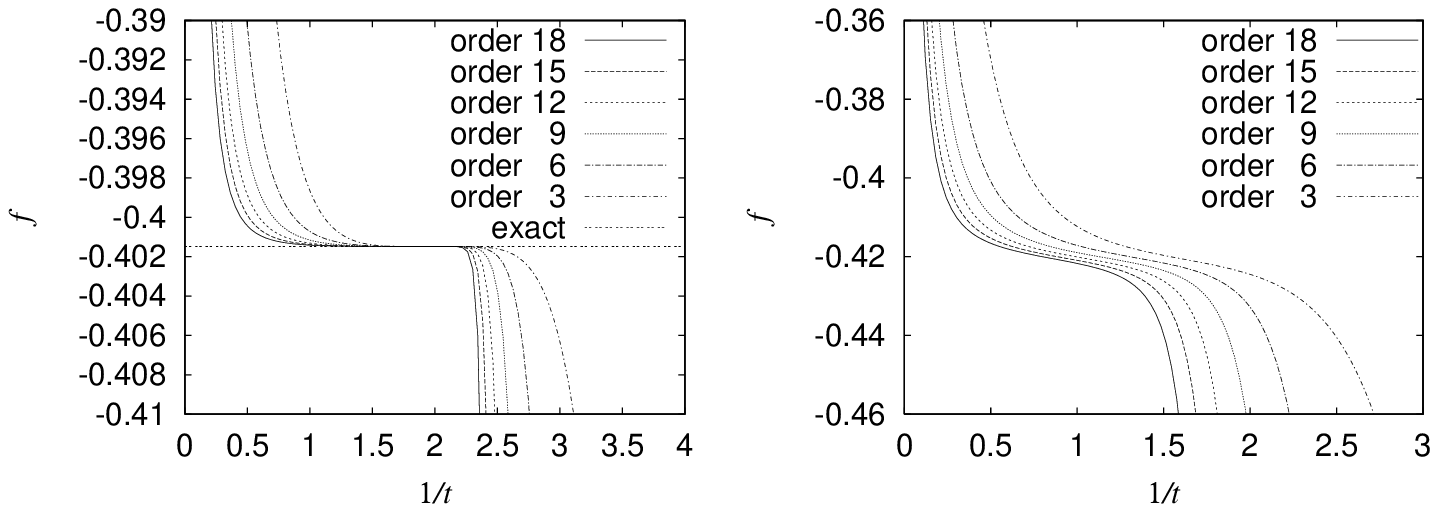,height=5cm} 
    \caption{{\bf (Left)} The free energy density $f$ 
is plotted as a function of $1/t$ for $g = -\frac{1}{12}$, $m = 1$
(critical point). Each curve corresponds to the order 3, 6, 9, 12, 15, 18.
The horizontal line represents the exact value
$f = 7/24 - \log 2 = -0.40148 \cdots$.
{\bf (Right)} The free energy density $f$ 
is plotted as a function of $1/t$ for $g = -\frac{1}{10}$, $m = 1$.
Each curve corresponds to the order 3, 6, 9, 12, 15, 18.
}
    \label{fig:fe_unbound}
}



In Fig.\ \ref{fig:fe_unbound} (right) 
we show the results for $g = -\frac{1}{10}$, $m = 1$ 
(the point C in Fig.\ \ref{fig:phase_diagram}), 
which is slightly below the critical point.
At each order we see a plateau-like region,
but the slope in that region increases with the order
(Note the striking difference from the figure on the left).
Thus the Gaussian expansion method seems to know the absence
of a stable vacuum with this choice of parameters.
This example also confirms the importance of the plateau formation
in the method. 

It is interesting to see how the plateau formation ceases to occur
when we cross the critical point.
This is reflected in the peculiar situation at the critical point.
The plateaus in Fig.\ \ref{fig:fe_unbound} (left)
do not have the tiny oscillations which become visible
upon magnification
in Fig.\ \ref{fig:fe_np_error} (left).
In fact in Fig.\ \ref{fig:fe_unbound} (left)
there is only one stationary point at each order 
we studied, and the free energy at that point
coincides with the exact value at the orders $\ge 2$.
The stationary point disappears as soon as we cross the critical point.



\subsubsection{Double-well potential ($g>0$, $m<0$)}
\label{double-well}

Let us then consider the case with the double-well potential.
The system undergoes a phase transition 
from the BIPZ phase to the two-lump phase
at $m= -2 \sqrt{g}$.
Fig.\ \ref{fig:fe_doublewell} (left) shows 
the free energy as a function of $1/t$
for $g = 1$, $m = -1$ (the point D in Fig.\ \ref{fig:phase_diagram}).
Although the chosen parameters are still well within the BIPZ phase,
we observe an oscillating behavior which becomes more and more violent
as the order is increased beyond the order 9.
This is also reflected in
Fig.\ \ref{fig:fe_doublewell} (right),
which shows 
the results extracted by the histogram technique at each order. 
The oscillation becomes even worse as we go to further negative $m$.

\FIGURE{\epsfig{file=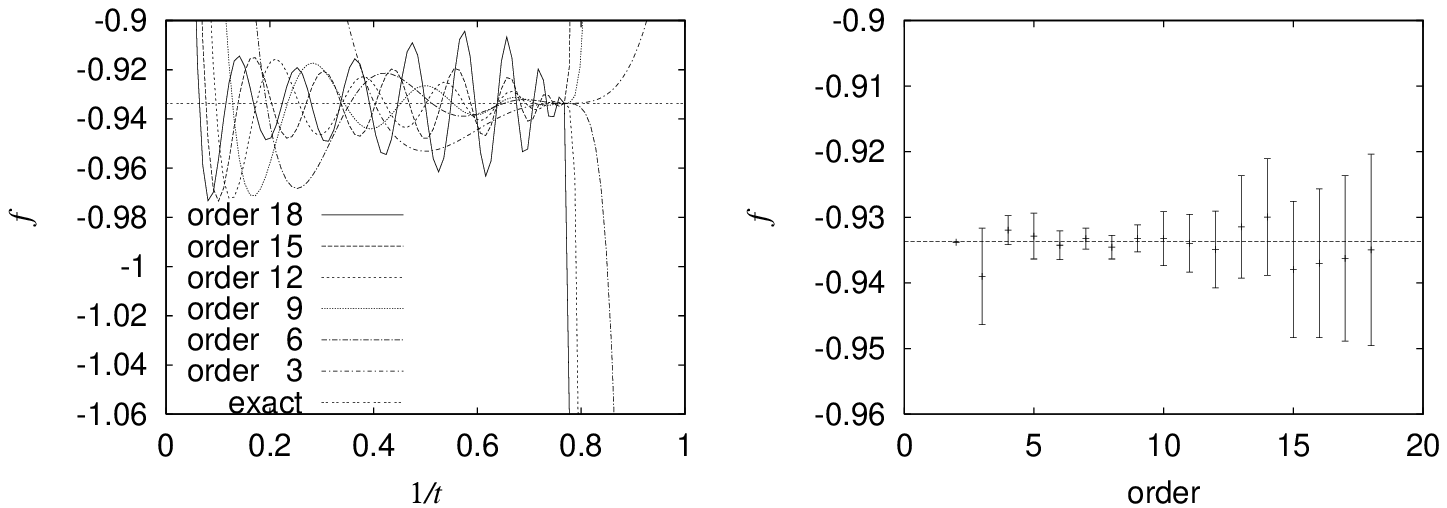,height=5cm}
    \caption{
{\bf (Left)} The free energy density $f$ 
is plotted as a function of $1/t$ for $g = 1$, $m = -1$.
Each curve corresponds to the order 3, 6, 9, 12, 15, 18.
The horizontal line represents the exact value.
{\bf (Right)} The free energy density $f$ obtained 
by the histogram prescription
at the order 2, ..., 18 for $g = 1$, $m = -1$.
(When we make a histogram, we used the region $0.01 \le  1/t \le  1$.)
The order 1 result ($f=-0.846574$) lies outside the displayed region.
The horizontal line represents the exact result
$f = -0.933705\cdots$.
}
    \label{fig:fe_doublewell}
}



Thus we find that the Gaussian expansion method in its simplest form
does not work in the case of the double-well potential.
(A similar result was obtained in the quantum mechanics
of a particle in a double-well potential \cite{conv}.)
This is understandable if we recall
that the method amounts to 
expanding the theory around $\phi = 0$,
which is a local maximum of the double-well potential.
We will reconsider the present example
in Section \ref{section:double-well-revisited} including a linear term
in the Gaussian action.



\setcounter{equation}{0}
\section{$\phi^3$ matrix model}
\label{section:linear}

In this Section we will test the Gaussian expansion method
in the $\phi^3$ matrix model, which is also exactly solvable.
In this model the potential is always unbounded from below,
but a stable vacuum exists in the large-$N$ limit when
the parameters in the potential are chosen appropriately.
The situation is thus similar to the $\phi^4$ model with
the unbounded potential, but the main issue here is that
we have to include a linear term in the Gaussian action
due to the absence of the Z$_2$ symmetry.
We will treat it in a way consistent with the loop expansion,
and check the result against the exact solution.
As a result of having a linear term, 
we have two free parameters, which makes
the identification of a plateau somewhat more nontrivial
than in the previous Section.
The histogram technique turns out to be quite useful.

\subsection{Exact solutions}
\label{exact_solutions2}

The $\phi^3$ matrix model is defined by the partition function
\beqa
Z & = & \int d^{N^2}\phi\, e^{-S}, \\
S & = & N\frac{1}{2} \,\tr \phi^2 + N\frac{g}{3} \,\tr \phi^3 \ , 
\label{phi3_action}
\eeqa  
where $\phi$ is a $N\times N$ hermitian matrix, and the measure
$d^{N^2}\phi$ is the same as in the $\phi^4$ matrix model.  
Although the action (\ref{phi3_action}) is unbounded from below, 
the theory is well-defined in the large-$N$ limit 
for $g^2 \leq g_c^2 \equiv \frac{1}{12\sqrt{3}}$.

The exact results for the free energy and observables 
are given by \cite{Brezin:1977sv} 
\beqa
F(g) & = & N^2\left[\frac12\ln N + \frac12\ln\frac{1+2\sigma}{2} 
-\frac13\frac{\sigma(3\sigma^2+6\sigma+2)}{(1+\sigma)(1+2\sigma)^2}
+O\left(\frac{\ln N}{N}\right)\right] \ , \\
\left\langle \frac1N \tr \phi \right\rangle & = & 
-g\frac{1+3\sigma}{(1+\sigma)(1+2\sigma)^2}  ~~~,~~~
\left\langle \frac1N \tr \phi^2 \right\rangle  =  
\frac{1+3\sigma}{(1+\sigma)(1+2\sigma)^2}  \ , 
\eeqa
where $\sigma$ is the largest solution of 
the equation $\sigma (1+\sigma)(1+2\sigma) = -2g^2$. 
The above expressions are valid for $|g| \leq g_c$. For $|g| > g_c$,
on the other hand,
the system does not have a stable vacuum even in the large-$N$ limit. 
In what follows we assume $g \ge 0$ without loss of generality
due to the duality under the sign flip of $g$.

\subsection{Systematic treatment of a linear term in the Gaussian action}

Let us denote the quadratic and cubic terms
in the action (\ref{phi3_action})
as $S_2$ and $S_3$ so that the action
reads $S=S_2 + S_3$.
Similarly to (\ref{GEM_action}), the Gaussian expansion
amounts to considering the action
\beq
S_{\rm GEM} (t, h; \lambda)
= \, \frac{1}{\lambda} \, \Bigl[ \Bigl\{ S_0 + \lambda (S_2 - S_0) 
\Bigr\}  + S_3  \Bigr] \  .
\label{GEM_action_phi3}
\eeq
Since the action (\ref{phi3_action}) does not have 
the Z$_2$ symmetry unlike the $\phi^4$ matrix model,
the $S_0$ in (\ref{GEM_action_phi3})
should include a linear term as
\beqa
 & & S_0 = S_{\rm G} + S_{\rm L}, \n
 & & S_{\rm G} = \frac{N}{2}t\,\tr \phi^2, \quad S_{\rm L} = Nh\,\tr \phi \ , 
\label{GSlinear}
\eeqa
where the real parameters $t$ and $h$ are arbitrary at this point.
Free energy and observables are calculated as 
an expansion with respect to $\lambda$,
and $\lambda$ is set to one eventually.
This procedure yields a loop expansion with 
the `classical action' $(S_0 + S_3)$
and the `one-loop counter term' $(S_2 - S_0)$.

As in the $\phi^4$ matrix model, let us rescale $\phi$ as
$\phi \mapsto \sqrt{\lambda}\phi$
so that the partition function takes the form
\beqa
Z &=& \int d^{N^2}\phi
\, e^{ - ( S_{\rm cl} + S_{\rm c.t.}) } \ ,  \\
S_{\rm cl} (t,h;\lambda) &=& S_{\rm G} 
+ \frac{1}{\sqrt{\lambda}} S_{\rm L}
+ \sqrt{\lambda} S_3   ~~~,~~~
S_{\rm c.t.} (t,h; \lambda) =  \lambda (S_2 - S_{\rm G})
- \sqrt{\lambda} S_{\rm L}  \ .
\label{Z_phi3_rescaled}
\eeqa
Thus the Gaussian expansion in the present case {\em cannot}
be viewed as an expansion with respect to $(S-S_0)$
unlike the $\phi^4$ matrix model.
The `counter terms' $S_{\rm c.t.}$ in (\ref{Z_phi3_rescaled})
can be easily implemented by using the relation
\beq
S_{\rm cl} (t,h ; \lambda)  + 
S_{\rm c.t.} (t,h ; \lambda) 
= S_{\rm cl} (t + \lambda(1-t) , h- \lambda h ; \lambda ) \ .
\label{ct_shift}
\eeq
Namely the calculation of a physical quantity
in the Gaussian expansion proceeds in two steps;
(i) obtain the $\lambda$-expansion of the quantity
using the `classical action' $S_{\rm cl} (t,h; \lambda)$,
(ii) shift the parameters $t$ and $h$ by $\lambda(1-t)$
and $(- \lambda h)$ respectively and reorganize the expansion
with respect to $\lambda$.

As usual we eliminate the linear term in $S_{\rm cl} (t,h;\lambda)$
by shifting the variables as 
$\phi = \tilde{\phi} + \frac{\alpha}{\sqrt{\lambda}}{\bf 1}$,
where the parameter $\alpha$ should satisfy
\beq
g\alpha^2 + t \alpha + h = 0 \ .
\label{quad_eq}
\eeq
The `classical action' then becomes
\beq
\label{action_after}
S_{\rm cl} (t,h;\lambda) 
 = C + N\frac{1}{2} \tilde{t} \,\tr \tilde{\phi}^2 
 + \sqrt{\lambda}N\frac{1}{3}g\,\tr \tilde{\phi}^3 \ ,
\eeq
where $C$ and $\tilde{t}$ are given by
\beqa
 C &=&  N^2 \frac{1}{\lambda}
   \left( \frac{1}{2}\alpha^2 t + h \alpha + \frac{1}{3}g \alpha^3
   \right) \ ,  \\
\tilde{t}  &=&  t +2g\alpha  \ .
\label{t_prime}
\eeqa
Thus the problem reduces to the conventional perturbative expansion 
in the $\phi^3$ matrix model.
Note that the constant term $C$ in (\ref{action_after})
contributes to the `order-$(-1)$ term' in the $\lambda$-expansion
of free energy.

\FIGURE{\epsfig{file=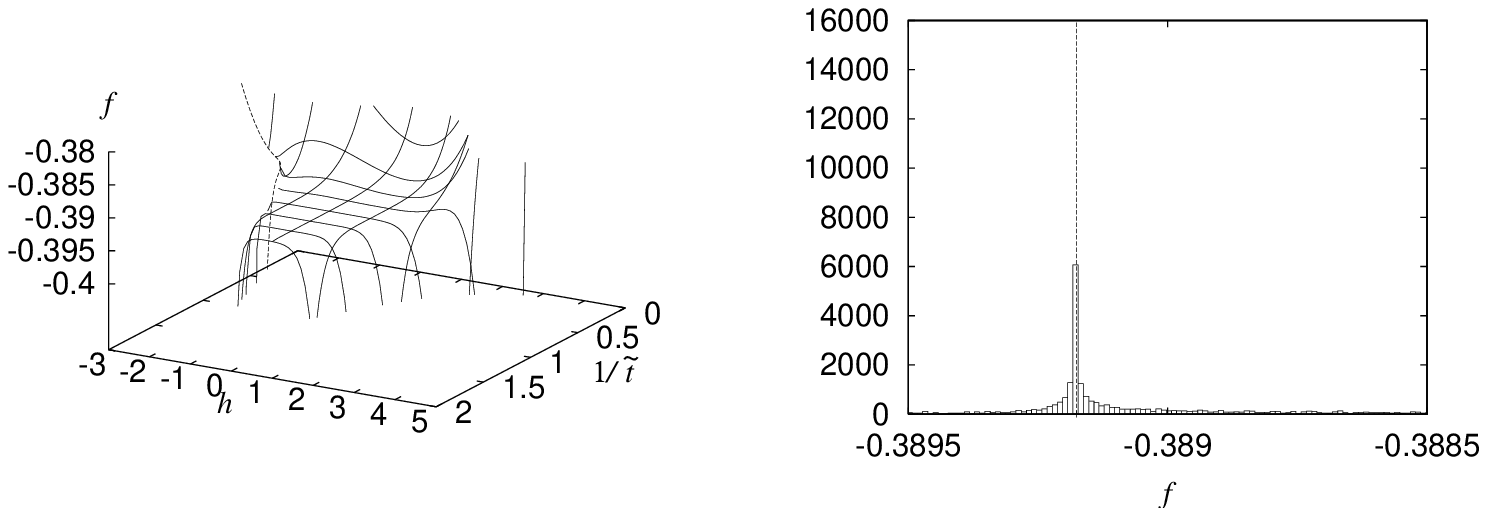,height=5cm}
    \caption{{\bf (Left)} Free energy density at the order 5
is plotted as a function of $1/\tilde{t}$ and $h$.
{\bf (Right)} The histogram of the free energy density.
(When we make a histogram, we used the region 
$0.02 \le  1/\tilde{t} \le  2$, $-3  \le  h \le  5$.
The bin size is taken to be $10^{-5}$.)
The vertical line represents the exact result $f=-0.389176\cdots$.}
    \label{fig:phi3_free5}
}

\FIGURE{\epsfig{file=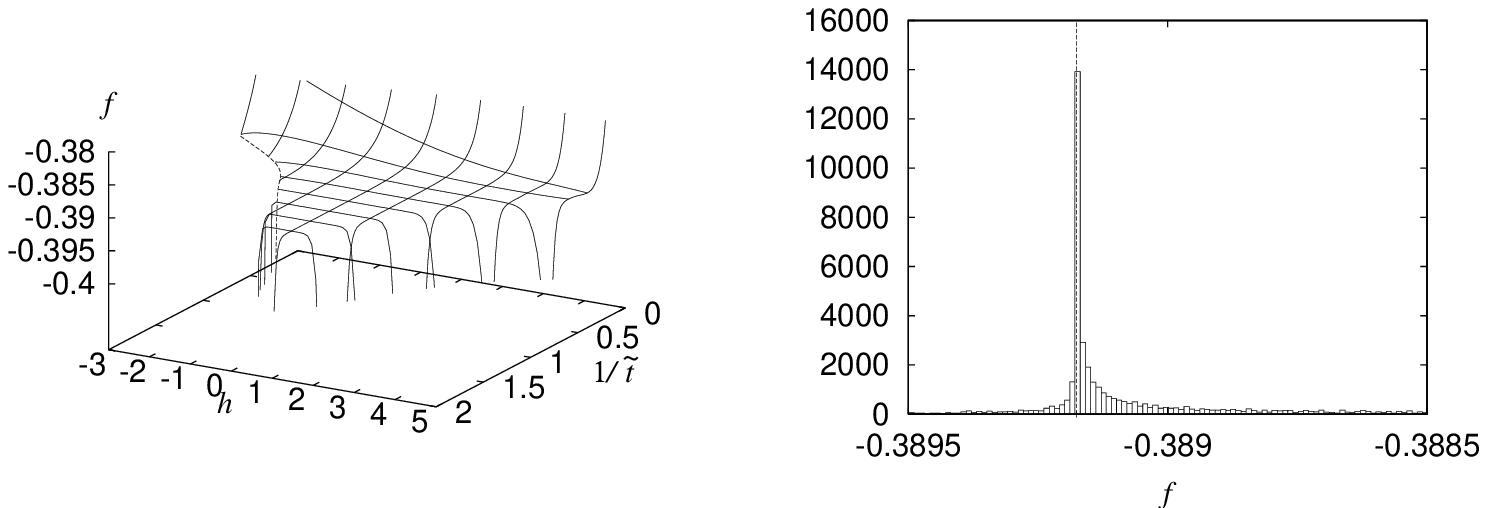,height=5cm}
    \caption{
The same as Fig.\ \ref{fig:phi3_free5} but for the order 15.
}
    \label{fig:phi3_free15}
}

Eq.\ (\ref{quad_eq}) has two solutions
$\alpha = \frac{1}{2g}(-t \pm \sqrt{t^2-4gh})$, 
which correspond to $\tilde{t} = \pm \sqrt{t^2-4gh}$ respectively.
Since $\tilde{t}$ is the coefficient of the quadratic term in eq.\
(\ref{action_after}), it should be positive in order for 
the Gaussian expansion to be well-defined.
This forces us to take
\beq
\alpha = \frac{1}{2g}(-t+\sqrt{t^2-4gh}) . 
\eeq
Considering that $1/\tilde{t}$ has the meaning as the
tree-level propagator,
we shall plot the result of the Gaussian expansion 
obtained for various $t$ and $h$
as a function of $1/\tilde{t}$ and $h$.
For each $\tilde{t}$ and $h$ satisfying $\tilde{t}^2 + 4 g h > 0$,
we obtain results corresponding to $t>0$ and $t<0$.
It turned out, however, that the plateau mainly develops in the branch
corresponding to $t>0$. We therefore plot the results for $t>0$ in 
what follows.

Here we show our results at the critical point
$g^2=g_c^2 \equiv \frac{1}{12\sqrt{3}}$.
In Fig.\ \ref{fig:phi3_free5} (left)
we plot the free energy density
defined by (\ref{fe_density})
as a function of $1/\tilde{t}$ and $h$ at the order 5.
We observe a clear plateau.
On the right we show the corresponding histogram,
which has a sharp peak 
at the exact value.
Fig.\ \ref{fig:phi3_free15} shows the results at the order 15.
The plateau extends and as a result the peak of the histogram becomes
higher.
(In the present case we were not able to obtain a reasonable estimate
for the uncertainty because the plateau turned out to be too flat.
This seems to be a peculiar feature at the critical point.)
We have checked that observables
such as $\langle \frac{1}{N}\tr\phi \rangle$
and $\langle \frac{1}{N} \tr\phi^2 \rangle$
also agree with the exact results very accurately.
These results support the correctness of the way we treated
the linear term in the Gaussian action.


\section{$\phi^4$ matrix model with the double-well potential, revisited}
\label{section:double-well-revisited}

In this Section we will revisit
the $\phi^4$ matrix model with the double-well potential,
where the simplest version of the Gaussian expansion seems to fail
(See Section \ref{double-well}).
Here we would like to perform the expansion including
a linear term as we formulated in the previous Section. 
In fact we are going to consider two kinds of linear term.
One is the same as in the $\phi^3$ matrix model,
and the other is a linear term with a `twist' 
which breaks the SU($N$) symmetry down to SU($N/2$)$\times$SU($N/2$).
For each case we obtain a plateau corresponding to
different large-$N$ solutions.
The first one corresponds to putting all the eigenvalues into one well,
and the second one corresponds to partitioning the eigenvalues equally
into the two wells. 
In accord with the exact results, the free energy is smaller for the latter,
which corresponds to the `true vacuum'.
The present example supports the validity of the strategy
adopted in the dynamical determination of the space-time dimensionality 
in the IIB matrix model \cite{Nishimura:2001sx,KKKMS,Kawai:2002ub}.



\subsection{Single-lump solution in the double-well potential}

In fact the $\phi^4$ matrix model with the double-well potential
has a one-parameter family of large-$N$ solutions corresponding to 
different partitioning of the $N$ eigenvalues into the two wells
(See Appendix \ref{sec:asym2cut}).
The solution (\ref{splitphase2}) corresponds to the equal partitioning,
and it has the smallest free energy.
For asymmetric partitioning, the free energy becomes larger due to 
the repulsive force and due also to the increase of the potential energy.
These asymmetric large-$N$ solutions are stable, however, against
tunneling of the eigenvalues because the potential barrier grows
linearly with $N$.
The situation is analogous to the existence of 
a stable large-$N$ solution in spite of the unbounded potential 
discussed in Section \ref{exact_solutions}.

Here we will consider the extreme case where all the eigenvalues are
put into one well.
The exact results can be readily obtained by applying the
technique used in Ref.\ \cite{Brezin:1977sv}.
The eigenvalue distribution has the form 
\beq
\rho(x) = \frac{1}{2\pi}\left\{ gx^2 + \sqrt{g\sigma}x -\frac23 (|m| -\sigma) 
\right \} \sqrt{(x - B_-)(B_+ - x)}
\label{excitation0} 
\eeq
for $ B_-  \le  x  \le  B_+ $ and $\rho(x)=0$ otherwise. 
The edges of the support are given by
\beq
B_{\pm} = \sqrt{\frac{\sigma}{g}} \pm \sqrt{\frac{2}{3g}(|m| - \sigma)} \ , 
\eeq
and the parameter $\sigma$ is defined as
\beq
\sigma = \frac{3|m| + 2\sqrt{m^2 - 15g}}{5} \ . 
\label{sigma-def-1cut}
\eeq
The free energy and observables are obtained as 
\begin{eqnarray}
  F &=& N^2\left[\frac12\ln N - \frac38
    + \frac12\ln\frac{5\sigma-|m|}{4} -\frac{43}{180}\frac{m^2}{g} \right.
  \n
&~& \left. \mbox{~~~~~~~}
 + \left(\frac53 - \frac2{45}\frac{m^2}{g}\right) \frac{|m|}{5\sigma-|m|}
     + O\left(\frac{\ln N}{N}\right) \right]  \ , 
\label{excitation1}\\
  \left\langle \frac1N \tr\phi \right\rangle &=& 4\sqrt{\frac{\sigma}{g}}
    \frac{\sigma}{5\sigma-|m|}  ~,~
  \left\langle \frac1N \tr\phi^2 \right\rangle = \frac{41}{45}\frac{|m|}{g}
    + \frac{16}{3}\left(\frac{1}{15}\frac{m^2}{g}-1\right)
    \frac{1}{5\sigma-|m|} \ .
\label{excitation2}
\end{eqnarray}
As can be seen from (\ref{sigma-def-1cut}),
the single-lump solution exists for $0 < g/m^2 \leq 1/15$,
while the Z$_2$-symmetric solution (\ref{splitphase2}) corresponding to 
the `true vacuum' exists for $0 < g/m^2 \leq 1/4$.
It is understandable that the critical $g/m^2$ is larger for the latter
(the potential wells become shallower as $g/m^2$ increases).


\subsection{Gaussian expansion method with a linear term}
\label{suN}

We will see that the single-lump solution (\ref{excitation0})
can be reproduced by the Gaussian expansion method
if we include a linear term.
The formulation is quite analogous to the $\phi^3$ matrix model,
so the description will be brief.
Instead of (\ref{GEM_action_phi3}) we consider the action
\beq
S_{\rm GEM} (t, h; \lambda)
= \, \frac{1}{\lambda} \, \Bigl[ \Bigl\{ S_0 + \lambda (S_2 - S_0) 
\Bigr\}  + S_4  \Bigr] \  ,
\label{GEM_action_phi4lin}
\eeq
where $S_0$ includes the linear term $S_{\rm L}$ as in (\ref{GSlinear}),
while $S_2$ and $S_4$ are the same as in Section \ref{basic_idea}.
Rescaling the variables as $\phi \mapsto \sqrt{\lambda}\, \phi$,
we obtain
\beqa
Z &=& \int d^{N^2}\phi
\, e^{ - ( S_{\rm cl} + S_{\rm c.t.}) } \ ,  \\
S_{\rm cl} (t,h;\lambda) &=& S_{\rm G} 
+ \frac{1}{\sqrt{\lambda}} S_{\rm L}
+ \lambda S_4   ~~~,~~~
S_{\rm c.t.} (t,h; \lambda) =  \lambda (S_2 - S_{\rm G})
- \sqrt{\lambda} S_{\rm L}  \ .
\label{Z_1cut_rescaled}
\eeqa
The `one-loop counter terms' can be incorporated by using
the relation
\beq
S_{\rm cl} (t,h ; \lambda)  + 
S_{\rm c.t.} (t,h ; \lambda) 
= S_{\rm cl} (t + \lambda(m-t) , h- \lambda h ; \lambda ) \ .
\label{ct_shift2}
\eeq
After eliminating the linear term in $S_{\rm cl} (t,h ; \lambda)$ 
by shifting the variable as
$\phi = \tilde{\phi} + \frac{\alpha}{\sqrt{\lambda}}{\bf 1}$,
the `classical action' becomes
\beq
\label{action_after2}
S_{\rm cl} (t,h;\lambda) 
 = C + N\frac{1}{2} \tilde{t} \,\tr \tilde{\phi}^2 
 + \sqrt{\lambda}N\frac{1}{3} \tilde{g} \,\tr \tilde{\phi}^3 
 + \lambda N \frac{1}{4}g\,\tr \tilde{\phi}^4 \ ,
\eeq
where $C$, $\tilde{t}$ and $\tilde{g}$ are given by
\beqa
C &=& 
N^2 \frac{1}{\lambda} \frac{\alpha}{4}(3h+t\alpha), \\
\tilde{t} & = & t +3g\alpha^2,  \\
\tilde{g} & = & 3g\alpha  \ .
\label{t_tilde}
\eeqa
Thus the problem reduces to the conventional perturbative 
expansion in the one-matrix model 
with both $\phi^3$ and $\phi^4$ interactions.
As in the previous cases, we can utilize the large-$N$ solution,
which can be obtained by the same method \cite{Brezin:1977sv}. 
See Appendix \ref{sec:mixed} for the details. 

The shift parameter $\alpha$ should satisfy
\beq
g \alpha^3 + t \alpha + h = 0 \ , 
\label{equation_alpha}
\eeq
which has three solutions. For $t>0$ they are given by
\beqa
\alpha_1 &  = & \frac{1}{18^{1/3}\sqrt{g}}\, 
\Bigl( - G_+^{1/3} + G_-^{1/3} \Bigr)  \ , \n
\alpha_2 & = & 
\frac{1}{18^{1/3}\sqrt{g}}\, 
\Bigl( e^{i\pi /3} G_+^{1/3} - e^{-i\pi /3} G_-^{1/3} \Bigr)  \ , \n
\alpha_3 & = & 
\frac{1}{18^{1/3}\sqrt{g}}\, 
\Bigl( e^{-i\pi /3} G_+^{1/3} - e^{i\pi /3} G_-^{1/3} \Bigr) \ ,  
\label{solution_alpha}
\eeqa
while for $t<0$ they are given by
\beqa
\alpha_1 &  = & \frac{1}{18^{1/3}\sqrt{g}}\, 
\Bigl( e^{-i\pi /3} G_+^{1/3} + G_-^{1/3} \Bigr)  \ , \n
\alpha_2 & = & 
\frac{1}{18^{1/3}\sqrt{g}}\, 
\Bigl( - G_+^{1/3} - e^{-i\pi /3} G_-^{1/3} \Bigr)  \ , \n
\alpha_3 & = & 
\frac{1}{18^{1/3}\sqrt{g}}\, e^{i\pi /3}
\Bigl( G_+^{1/3} -  G_-^{1/3} \Bigr)  \ ,
\label{solution_alpha2}
\eeqa
where we have introduced
$G_{\pm} \equiv  \pm 9\sqrt{g}h + \sqrt{81gh^2+ 12 t ^3}$.
Among the three branches, two of them ($\alpha_2$ and $\alpha_3$ for $t>0$, 
$\alpha_1$ and $\alpha_2$ for $t<0$) are mapped to each other 
under the sign flip of $h$, while the remaining one is symmetric. 
This structure is a consequence of the Z$_2$ symmetry of the
$\phi^4$ model. 
Requiring $\tilde{t}$ to be real positive
%
we are left with the following four cases.
\beq
\begin{array}{ll}
\alpha = \alpha_1 
& \quad \mbox{for }\quad  t > 0 \ , \\
\alpha =  \alpha_1 
& \quad \mbox{for }\quad  t < 0, \quad 
h  \leq -\sqrt{\frac{4}{27g}} |t|^{3/2}
\ , 
\\
\alpha =\alpha_2 
& \quad \mbox{for }\quad  t < 0, \quad  
h  \geq \sqrt{\frac{4}{27g}} |t|^{3/2}
\ , 
\\
\alpha = \alpha_1, \alpha_2
& \quad \mbox{for }\quad  t < 0, \quad 
-\sqrt{\frac{4}{27g}} |t|^{3/2} < 
h  <  \sqrt{\frac{4}{27g}}  |t|^{3/2} \ . 
\end{array}
\label{region_excited}
\eeq
As in the $\phi^3$ model, we plot the result of the 
Gaussian expansion as a function of $1/\tilde{t}$ and $h$.
In what follows we show only the branch which includes the plateau
in the $h>0$ regime.


We present the result for the free energy at $g=1/15$ and $m=-1$
(the point E in Fig.\ \ref{fig:phase_diagram}),
which is the critical point for the existence of the single-lump solution.
In Fig.\ \ref{fig:order3} (left)
the free energy density $f$ at the order 3 
is plotted as a function of $1/\tilde{t}$ and $h$.
The figure on the right shows the corresponding histogram.
(Here again we were not able to obtain a reasonable estimate
for the uncertainty because the plateau turned out to be too flat.)
The formation of a plateau is clearly observed
and the histogram has a sharp peak at the exact result.
Fig.\ \ref{fig:order6} shows the results at the order 6.
The plateau extends to larger area, and consequently
the peak of the histogram becomes higher.

\FIGURE{\epsfig{file=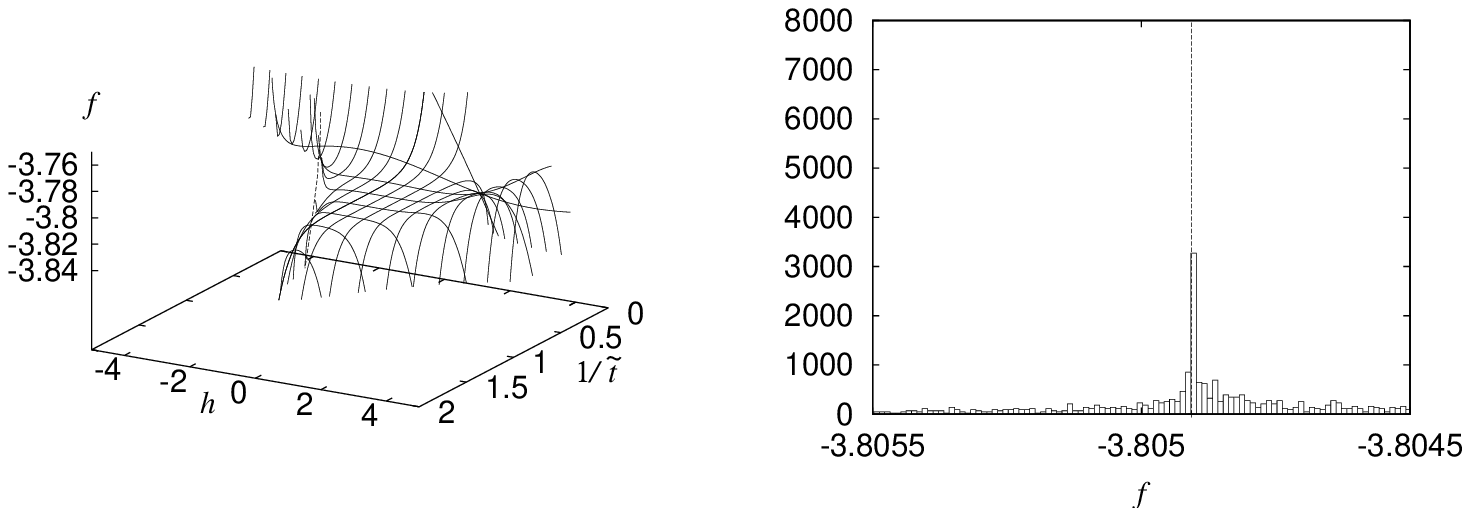,height=5cm} 
    \caption{
{\bf (Left)} Free energy density at the order 3
is plotted as a function of $1/\tilde{t}$ and $h$ 
for $g = 1/15$ and $m = -1$.
{\bf (Right)} The histogram of the free energy density.
(When we make a histogram, we used the region 
$0.02 \le  1/\tilde{t} \le  2$, $-5  \le  h \le  5$.
The bin size is taken to be $10^{-5}$.)
The vertical line represents the exact result  $f = -3.80491\cdots$.
}
    \label{fig:order3}
}


\FIGURE{\epsfig{file=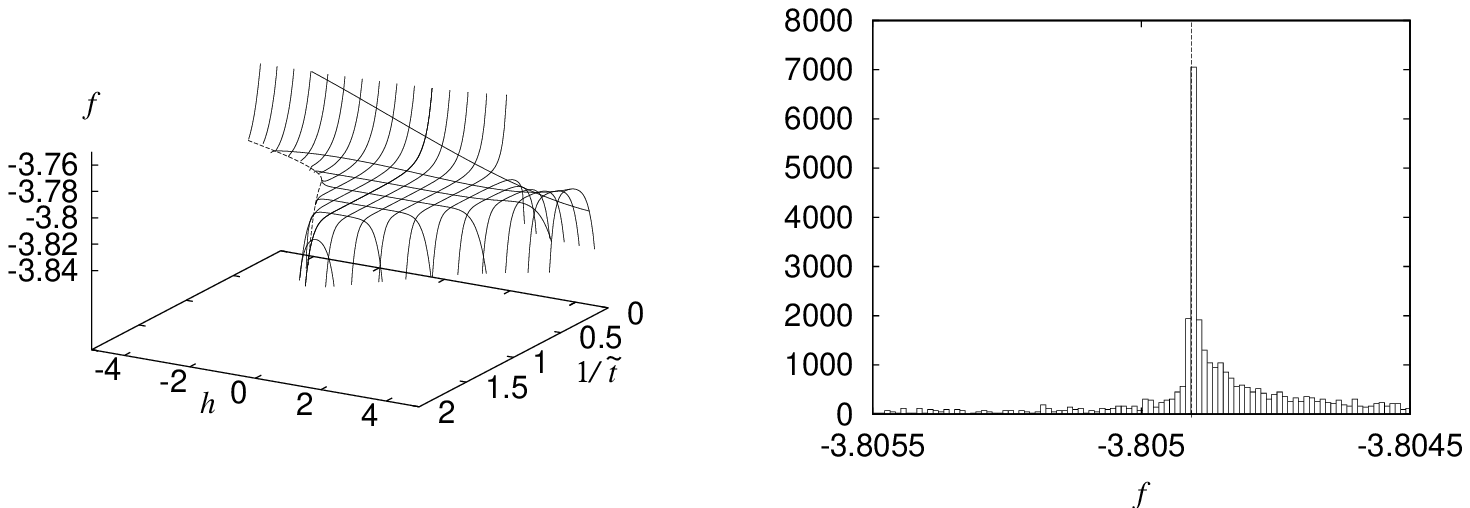,height=5cm} 
    \caption{
The same as Fig.\ \ref{fig:order3} but for the order 6.
}
    \label{fig:order6}
}


\subsection{Gaussian expansion method with a twisted linear term}
\label{vacuum-split}

In this Section we attempt to reproduce the `true vacuum', 
which corresponds to partitioning the eigenvalues equally into the two wells.
The failure of the Gaussian expansion method in Section
\ref{double-well} may be attributed to the fact that the 
$\lambda$-expansion with the action (\ref{GEM_action})
corresponds to expanding around the classical solution $\phi=0$,
where the double-well potential becomes locally maximum.
(When we say `classical', we regard the parameter $\lambda$ as 
the Planck constant $\hbar$.)
Here we will take the Gaussian term $S_0$
in the action (\ref{GEM_action})
in such a way that the classical solution becomes
proportional to 
\beq
J = \sigma_3 \otimes \id_{N/2} 
= {\rm diag}(\underbrace{1,\cdots, 1}_{N/2}, 
\underbrace{-1, \cdots, -1}_{N/2}) \ ,
\label{defJ}
\eeq
where and henceforth $\sigma_a$ denote Pauli matrices.
This can be achieved by considering the action 
(\ref{GEM_action_phi4lin}) with a `twisted linear term'
given by
\beq
S_{\rm L} = Nh\,\tr (J\phi) \ .
\label{GSlinear2}
\eeq
This term breaks the SU($N$) symmetry
down to SU($N/2$)$\times$SU($N/2$), but the 
total action (\ref{GEM_action_phi4lin}) has the `Z$_2$ symmetry'
\beq
\phi \mapsto - T \phi T^\dag ~~~,~~~
T = \sigma_1 \otimes \id_{N/2} \ .
\eeq
We expect that other large-$N$ solutions described in Appendix
\ref{sec:asym2cut} can be reproduced by considering the asymmetric twist
\beq
J' = {\rm diag}(\underbrace{1,\cdots, 1}_{n_{\rm L}}, 
\underbrace{-1, \cdots, -1}_{n_{\rm R}})~~~,~~~n_{\rm L} + n_{\rm R} = N
\eeq
instead of (\ref{defJ}).

After eliminating the linear term by shifting the variable as
$\phi = \tilde{\phi} + \frac{\alpha}{\sqrt{\lambda}} J$,
the `classical action' becomes
\beq
\label{action_after3}
S_{\rm cl} (t,h;\lambda) 
 = C + N\frac{1}{2} \tilde{t} \,\tr \tilde{\phi}^2 
    + N\frac12 g\, \alpha^2\, \tr(J\phi J\phi)
    + \sqrt{\lambda}N\tilde{g}\,\tr (J\phi^3)
 + \lambda N \frac{1}{4}g\,\tr \tilde{\phi}^4 \ ,
\eeq
where $C$, $\tilde{t}$ and $\tilde{g}$ are given by
\beqa
C &=& 
N^2 \frac{1}{\lambda} \frac{\alpha}{4}(3h+t\alpha), \\
\tilde{t} & = & t +2g\, \alpha^2,  \\
\tilde{g} & = & g\, \alpha  \ .
\label{t_tilde2}
\eeqa
By decomposing the matrix $\phi$ as 
\beq
\phi = \sum _{a=1}^{4} \sigma_a \otimes \varphi_a \ ,
\eeq
where $\sigma_4$ is a $2\times 2$ unit matrix and $\varphi_a$
are $N/2 \times N/2$ hermitian matrices,
the quadratic term in  (\ref{action_after3}) becomes
\beqa
S_{\rm quad} &=& N ( \tilde{t} - g \, \alpha^2) \, 
\tr \{  (\varphi_1)^2 +(\varphi_2)^2 \} \n
&~& + N ( \tilde{t} + g \, \alpha^2)  \, 
\tr \{ (\varphi_3)^2 +(\varphi_4)^2 \} \ .
\label{action_after4}
\eeqa
In order to make the Gaussian expansion well-defined,
we therefore have to require $\tilde{t} - g\, \alpha^2 >0$, which means 
\beq
t + g\, \alpha^2 > 0 \ . 
\label{morecondition}
\eeq

\FIGURE{\epsfig{file=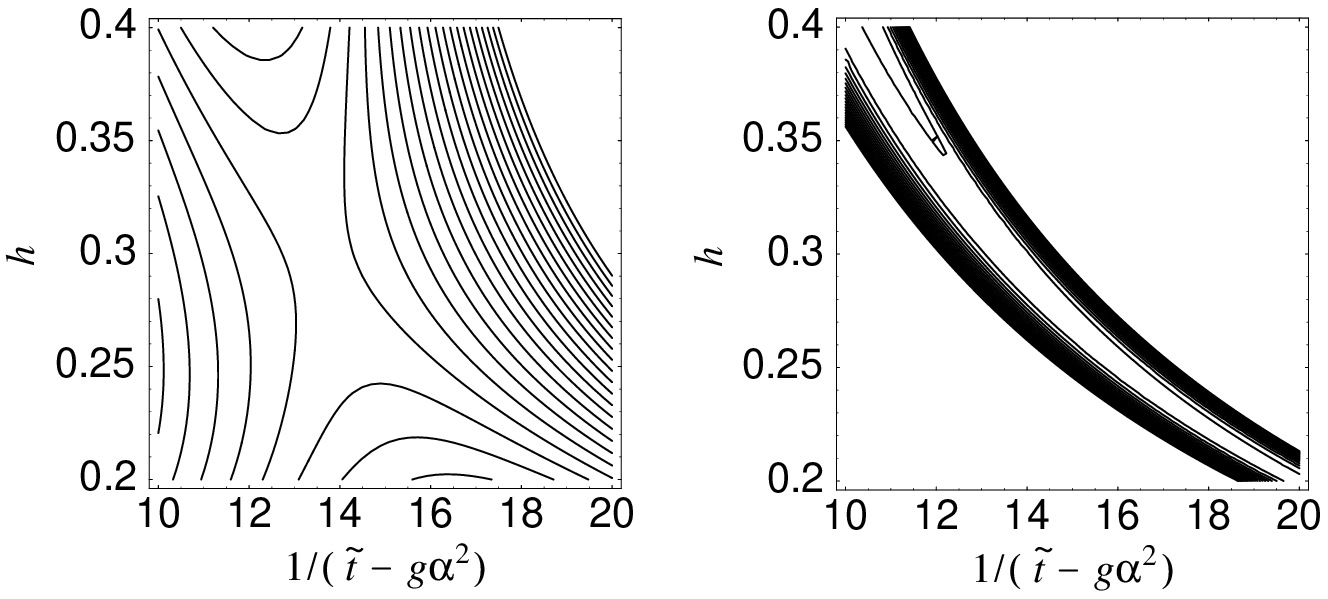,height=5cm}
    \caption{{\bf (Left)} 
A contour plot of the free energy density 
as a function $1/(\tilde{t}-g\alpha^2)$ and $h$ at the order 1 
for $m = -1$, $g = \frac{1}{15}$. 
40 contours for $f = -5.5 + 0.02\times n$ $(n = 1,\cdots, 40)$ 
are plotted. The blank region near the right-up corner means that
the calculated free energy density does not fall 
in the specified region of $f$.
{\bf (Right)} 
The same as in the figure on the left but for the order 3.
The plateau has grown diagonally from the left-top to the right-bottom. 
The meaning of the blank regions is the same as in the left figure.
}
    \label{fig:toukousen}
}

The equation for $\alpha$ is the same as (\ref{equation_alpha}),
so the solutions are (\ref{solution_alpha}), (\ref{solution_alpha2}). 
Combining (\ref{morecondition}) and (\ref{equation_alpha}),
one finds that $\alpha$ and $h$ should have opposite signs.
Thus the allowed real solution for $\alpha$ is determined uniquely 
in the present case as 
\beq
\begin{array}{ll}
\alpha = \alpha_1 & \quad \mbox{for }\quad t >0 \\
\alpha = \alpha_1 & \quad \mbox{for }\quad t < 0, \, h < 0 \\
\alpha = \alpha_2 & \quad \mbox{for }\quad t < 0, \, h > 0 \ .
\end{array}
\eeq
Note that $t$ and $h$ can take any real values 
except for $t<0$ and $h=0$.
As before we plot the results obtained for various $t$ and $h$ 
as a function of $1/(\tilde{t}- g \alpha^2)$ and $h$.
In the present case we have only one branch.
Since the figure is invariant under the reflection $h \mapsto -h$
due to the Z$_2$ symmetry,
we will restrict ourselves to $h>0$ in what follows.

Since the one-matrix model with the
SU($N$) breaking terms 
(\ref{action_after3}) cannot be solved exactly with the known techniques,
we have to perform diagrammatic calculations explicitly
unlike the previous cases
(See Appendix \ref{sec:SDmethod} for the details).
We have performed the calculations up to the order 3.
In Fig.\ \ref{fig:toukousen} we show 
contour plots of the free energy density 
as a function of $1/(\tilde{t}- g \alpha^2)$ and $h$
at the order 1 and 3 for $m = -1$, $g = \frac{1}{15}$ 
(the point E in Fig.\ \ref{fig:phase_diagram};
the same as in Section \ref{suN}). 
\TABLE[pos]{
\\
\\
\begin{tabular}{|c|l|l|l|}
\hline \hline
\multicolumn{1}{|c|}{quantity}   & 
\multicolumn{1}{c|}{$f$}   & 
\multicolumn{1}{c|}{$\frac{1}{N}\langle \tr \phi^2 \rangle$}   & 
\multicolumn{1}{c|}{$\frac{1}{N}\langle \tr \phi^4 \rangle$}    \\
\hline
order 1    & $-4.82843$  & $15.8906$ &  $241.361$ \\
order 3    & $-4.82861$   & $15.8899$ & $241.346$   \\
\hline
exact    & $-5.14859$   & $ 15$ & $ 240$   \\
\hline \hline
\end{tabular}
  \caption{The results of the free energy and the observables
at the order 1 and 3.
}
\label{tab:fe_trH}
}

Since our analysis here is restricted to low orders,
where the use of the histogram method is not highly motivated,
we evaluated the free energy at the extrema
on the plateau as in Refs.\ \cite{Nishimura:2001sx,KKKMS,Kawai:2002ub}.
Although there are more than one extrema at the order 3,
the difference in the results is negligible within the precision
we are discussing.
The observables 
$ \langle \frac1N\tr \phi^2\rangle$ and $\langle \frac1N\tr \phi^4\rangle$ 
are also evaluated at the extrema of the free energy
as in Refs.\ \cite{Nishimura:2001sx,KKKMS,Kawai:2002ub}.
The results are shown in Table \ref{tab:fe_trH}.
We see a reasonable agreement with the exact results already at
the order 1. The trend of improvement seen at the order 3, however,
turns out to be tiny. This might be related to the fact that 
the twisted linear term breaks the full SU($N$) symmetry,
which is expected to be restored at higher orders.

Let us emphasize that the results 
in Sections \ref{suN} and \ref{vacuum-split}
are obtained for the same action with the same parameters.
Depending on the choice of the Gaussian action, we have seen
plateaus with different free energy.
Each plateau corresponds to a large-$N$ saddle-point solution,
and the true vacuum can be identified with the plateau which has 
the smallest free energy.
Thus our results support the strategy adopted in the dynamical 
determination of the space-time dimensionality
in the IIB matrix model \cite{Nishimura:2001sx,KKKMS,Kawai:2002ub}.

\section{Summary and discussions}
\label{section:summary}

In this paper we have tested the Gaussian expansion method
in the one-matrix model, which is exactly solvable and is well-understood
in the context of 2d quantum gravity and non-critical string theory.

One of the most important findings is that
the method is able to reproduce large-$N$ solutions which are not 
stable at finite $N$, but stabilize only in the large-$N$ limit.
Examples we have seen are the vacuum in the unbounded potential
(as in the $\phi^4$ model with $g<0$, $m>0$ and the $\phi^3$ model)
and the single-lump solution in the double-well potential
($\phi^4$ model with $g>0$, $m<0$).
This enables us to understand the situation encountered
in the study of the IIB matrix model, where we obtained many plateaus with
different space-time dimensionality.
We may naturally expect that each plateau corresponds to 
some large-$N$ saddle-point solution
and that the true vacuum can be identified with 
the plateau which gives the smallest free energy
as in the case of the double-well potential.
In fact this was assumed implicitly
in the dynamical determination of the space-time dimensionality,
but now we have provided a concrete
example where the statement holds.

On the technical side, 
we would like to emphasize the importance of 
the plateau formation in this method.
We have observed in the case of the unbounded potential
that the plateau formation is quite sensitive to the
stability of the vacuum.
We have also formulated a prescription to
include a linear term, which was shown to work in various examples.
In these cases we have to deal with two free parameters, 
which makes the identification of a plateau more nontrivial. 
Here we find the histogram technique to be quite useful.

To conclude, we hope our results clarified the nature of the 
Gaussian expansion method and confirmed its usefulness 
in particular in matrix model applications.
%
We expect that the method can still be refined or extended 
in many different ways,
thus allowing us to study various systems which are not easily
accessible by other means.

\acknowledgments
We thank K.~Nishiyama for his participation at the earlier stage of
this work.
The work of J.N.\ is supported in part by Grant-in-Aid for 
Scientific Research (No.\ 14740163) from 
the Ministry of Education, Culture, Sports, Science and Technology.


\appendix

\section{General solutions in the double-well potential}
\label{sec:asym2cut}
\setcounter{equation}{0}
\renewcommand{\theequation}{A.\arabic{equation}}

In this Appendix we discuss the general 
large-$N$ solutions in the double-well potential 
($m<0$, $g>0$ case of the $\phi^4$ matrix model).
Each solution corresponds to different partitioning of 
the $N$ eigenvalues into the two wells, say,
$n_{\rm L}$ in the left well and
$n_{\rm R}$ in the right well, 
where $n_{\rm L} + n_{\rm R} = N$. 
In the large-$N$ limit, 
the support of the eigenvalue distribution
is given by 
$L \equiv [-B', -A'] \cup [A, B]$.
As a parameter which characterizes each solution, we introduce
\beq
\nu \equiv \left (\frac{A' - B'}{A- B}\right )^2 \ , 
\eeq
which takes the values $\nu = 0,1,\infty$ corresponding to 
$(n_{\rm L}, n_{\rm R}) = (0,N)$, $(N/2, N/2)$, $(N, 0)$, respectively. 
Since the solution for $1/\nu$ is Z$_2$-equivalent
to that for $\nu$, we may restrict ourselves to $\nu \le 1$.
Introducing the variables 
\beq
s \equiv \sqrt{g}\, \frac{A+B}{2} ~~~,~~~
s' \equiv \sqrt{g} \, \frac{A'+B'}{2} \ ,
\eeq
the saddle-point condition reads
\beqa
 & & (s-\nu s')(|m|-s^2-s'^2+ss')+ \frac{1+\nu}{2}ss' (s-s') = 0,   
\n
& & \left(\frac{1-\nu}{2}\right)^2 (|m|-s^2-s'^2+ss')^2 \n
& & \quad + 
\frac{1+\nu}{2}(|m|-s^2-s'^2+ss')(s+s')\{(2-\nu)s+(2\nu -1)s'\} 
= g(1+\nu)^2 \ ,
\label{saddle}
\eeqa
which determines $s$ and $s'$ for any $\nu$,
and the eigenvalue distribution has the form 
\beq
\rho(x) = \left \{ \begin{array}{ll}
\frac{1}{2\pi}\left\{ gx+\sqrt{g}(s-s')\right\}\sqrt{(x-A)(B-x)(x+A')(x+B')}
& \mbox{ for } x\in [A, B] \\
\frac{1}{2\pi}\left\{ -gx-\sqrt{g}(s-s')\right\}\sqrt{(A-x)(B-x)(-x-A')(x+B')}
& \mbox{ for } x\in [-B', -A']   \ .
\end{array}
\right. 
\eeq
The equations (\ref{saddle}) are not solvable analytically
for general $\nu$ except for $\nu=0, 1$,
at which one obtains the analytic solutions 
(\ref{splitphase2})
and 
(\ref{excitation0}) respectively.     

\section{Exact solution for the model with both
$\phi^3$ and $\phi^4$ interactions}
\label{sec:mixed}
\setcounter{equation}{0}
\renewcommand{\theequation}{B.\arabic{equation}}

In Section \ref{suN} the problem reduced to 
the $\lambda$-expansion of the free energy ${\cal F}$ defined by
\beqa
\exp[-N^2 {\cal F}] & \equiv & 
\int d^{N^2}\phi \, \exp\left[-N \frac12 \tilde{t} \,\tr \phi^2 
-\sqrt{\lambda}N\frac13\tilde{g}\,\tr \phi^3 -\lambda N\frac14 g\,\tr \phi^4
\right] \ .
\label{mixedZ} 
\eeqa
The one-matrix model with both $\phi^3$ and $\phi^4$ interactions
can be solved in the large-$N$ limit using the method of
Ref.\ \cite{Brezin:1977sv}. 
Assuming that the eigenvalue distribution $\rho (x)$ defined by 
(\ref{defrho}) has a finite support $[2a, 2b]$ in the large-$N$ limit,
it takes the form
\beqa
\rho(x) & = & \frac{1}{2\pi}(\lambda g x^2 - A x -B) \sqrt{(x-2a)(2b-x)} \ , \\
A & \equiv & -\lambda g (a +b) -\sqrt{\lambda}\tilde{g} \ , \\
B & \equiv & -\tilde{t} -\frac12\lambda g (b-a)^2 -\lambda g (a+b)^2 
-\sqrt{\lambda} \tilde{g} (a + b) \ . 
\eeqa
In order to simplify the expressions, we introduce
$$
 g=\tilde{t}^2\kappa \ , \quad 
\tilde{g}=\tilde{t}^{\frac32}\tilde{\kappa} \ , \quad 
a+b=\sqrt{\frac{\lambda}{\tilde{t}}}\tilde{\kappa}s \ , \quad 
(b-a)^2 = \frac{1}{\tilde{t}} \Delta  \ ,
$$ 
where the parameters $s$ and $\Delta$ are determined by 
\beqa
\Delta & = & \frac{-2s}{1+3\lambda\kappa s}
(1 + \lambda \tilde{\kappa}^2 s + \lambda^2 \kappa\tilde{\kappa}^2 s^2) \ , 
\label{mixed_b-a} \\
s & = & \frac{-2}{1+12\lambda \kappa}\left[1+
\frac34 \lambda (\kappa + 2\tilde{\kappa}^2 + 12\lambda \kappa^2)s^2 
+\lambda^2 (5\kappa + \tilde{\kappa}^2)\tilde{\kappa}^2 s^3 \right. \n 
 & & \qquad \left. + \frac14 \lambda^3 (18\kappa + 19\tilde{\kappa}^2) 
\kappa \tilde{\kappa}^2s^4 +\frac{15}{2}\lambda^4\kappa^2\tilde{\kappa}^4 s^5 
+ \frac{15}{4} \lambda^5 \kappa^3 \tilde{\kappa}^4 s^6 \right] \ . 
\label{mixed_s}
\eeqa
In terms of $s$ and $\Delta$, the free energy is given as 
\beqa
{\cal F} & = & \frac12 \ln N -\frac54 + \frac12 \ln \frac{2\tilde{t}}{\Delta} 
+ \frac12\lambda\tilde{\kappa}^2s^2 + \frac13\lambda^2\tilde{\kappa}^4s^3  
+\frac14\lambda^3\kappa\tilde{\kappa}^4s^4 
+\frac{5}{16}(1 + 2\lambda\tilde{\kappa}^2s + 
3\lambda^2\kappa\tilde{\kappa}^2s^2)\Delta  \n 
 & & + \frac{5}{32}\lambda\kappa\Delta^2 
+\frac{1}{32}\lambda\tilde{\kappa}^2s(1+\lambda\tilde{\kappa}^2s 
+\lambda^2\kappa\tilde{\kappa}^2s^2)(1+3\lambda\kappa s)\Delta^2 \n 
 & & +\frac{1}{256}\lambda \left(\kappa 
+\frac23\tilde{\kappa}^2 (2+9\lambda\kappa s) 
+\lambda\kappa\tilde{\kappa}^2 s(4+15\lambda\kappa s)\right) \Delta^3  
+ \frac{3}{2048}\lambda^2\kappa^2\Delta^4 \n 
 & & + O\left(\frac{\ln N}{N}\right) \ .
\eeqa
Although we cannot solve (\ref{mixed_s}) with respect to $s$  
in a closed form, 
we can obtain $s$ order by order in $\lambda$ by iteration,
which is sufficient for our purpose.
Up to O$(\lambda^6)$, the parameter $s$ and the free energy
${\cal F}$ are given respectively as
\beqa
s & = & -2 + 6\lambda (3\kappa -2\tilde{\kappa}^2) 
-4\lambda^2 (45\kappa^2 -92\kappa\tilde{\kappa}^2 +32\tilde{\kappa}^4) \n
 & & +10\lambda^3 (189\kappa^3 -786\kappa^2\tilde{\kappa}^2 
+712\kappa\tilde{\kappa}^4 -168\tilde{\kappa}^6) \n 
 & & -12\lambda^4 (1701\kappa^4 -11952\kappa^3\tilde{\kappa}^2 
+20064\kappa^2\tilde{\kappa}^4 -11464\kappa\tilde{\kappa}^6 
+ 2048 \tilde{\kappa}^8) \n 
 & & +28\lambda^5 (8019\kappa^5 -85482\kappa^4\tilde{\kappa}^2 
+ 228240\kappa^3\tilde{\kappa}^4 -230032\kappa^2\tilde{\kappa}^6 
+ 95488\kappa\tilde{\kappa}^8 -13728\tilde{\kappa}^{10}) \n 
 & & - 8\lambda^6 (312741\kappa^6 -4712580\kappa^5\tilde{\kappa}^2 
+18297792 \kappa^4\tilde{\kappa}^4 - 28301808\kappa^3\tilde{\kappa}^6 
+20097664 \kappa^2\tilde{\kappa}^8  \n 
 & & \qquad -6537600\kappa\tilde{\kappa}^{10} 
+786432\tilde{\kappa}^{12}) + O(\lambda^7) \ , 
\\
{\cal F} & = & \frac12\ln N + \frac12\ln \frac{\tilde{t}}{2} 
+ \frac16\lambda (3\kappa -4\tilde{\kappa}^2) 
+\lambda^2 \left(-\frac98\kappa^2 +6\kappa\tilde{\kappa}^2 
-\frac83\tilde{\kappa}^4\right) \n
 & & +\lambda^3 \left(\frac92\kappa^3 -54\kappa^2\tilde{\kappa}^2 
+68\kappa\tilde{\kappa}^4 -\frac{56}{3}\tilde{\kappa}^6\right) \n 
 & & +\lambda^4 \left(-\frac{189}{8}\kappa^4 +504\kappa^3\tilde{\kappa}^2 
-1236\kappa^2\tilde{\kappa}^4 + 856\kappa\tilde{\kappa}^6 
-\frac{512}{3}\tilde{\kappa}^8\right) \n 
 & & +\lambda^5 \left( \frac{729}{5}\kappa^5 -4860\kappa^4\tilde{\kappa}^2 
+ 19584\kappa^3\tilde{\kappa}^4 -24608\kappa^2\tilde{\kappa}^6 
+ 11680\kappa\tilde{\kappa}^8 -\frac{9152}{5}\tilde{\kappa}^{10}\right) \n 
 & & +\lambda^6 \left(-\frac{8019}{8}\kappa^6 +48114\kappa^5\tilde{\kappa}^2 
-288468\kappa^4\tilde{\kappa}^4 + 567816\kappa^3\tilde{\kappa}^6 
-469472\kappa^2\tilde{\kappa}^8 \right. \n 
 & & \qquad \left. +169152\kappa\tilde{\kappa}^{10} 
-\frac{65536}{3}\tilde{\kappa}^{12} \right) +O(\lambda^7) \ .
\eeqa

\section{Diagrammatic calculations with a twisted linear term}
\label{sec:SDmethod}
\setcounter{equation}{0}
\renewcommand{\theequation}{C.\arabic{equation}}

In this Appendix
we describe the diagrammatic calculations in the presence
of the twisted linear term.
As we discussed in Section \ref{vacuum-split},
the problem reduces 
to the $\lambda$-expansion of the free energy 
${\cal F}(\mu ,r)$ defined by
\beqa
\exp[-N^2 {\cal F}(\mu,r)] & \equiv & 
\int d^{N^2}\phi \, \exp\left[-N \frac12 \mu \,\tr \phi^2 
-N\frac12 r \,\tr (J\phi J\phi) \right. \n
 & &   \left. \hspace{2cm}
-\sqrt{\lambda}N\tilde{g}\,\tr (J\phi^3) -\lambda N\frac14 g\,\tr \phi^4
\right] \ ,
\label{ZJphi34B} 
\eeqa
where the parameters $\mu$ and $r$ are related to those
in (\ref{action_after3}) as $\mu = \tilde{t}$ and $r = g\alpha^2$.
The Feynman rules are given in 
Fig.\ \ref{fig:feynman}. 
Instead of evaluating all the diagrams contributing to ${\cal F}(\mu,r)$,
we use the Schwinger-Dyson equations to
reduce the number of diagrams to be evaluated.
Here we extend the idea of Ref.\ \cite{KKKMS} to the case including
a linear term. 
In what follows we restrict ourselves to the large-$N$ limit,
so that we have to consider planar diagrams only,
but the method itself is applicable to finite $N$ as well.

\FIGURE{\epsfig{file=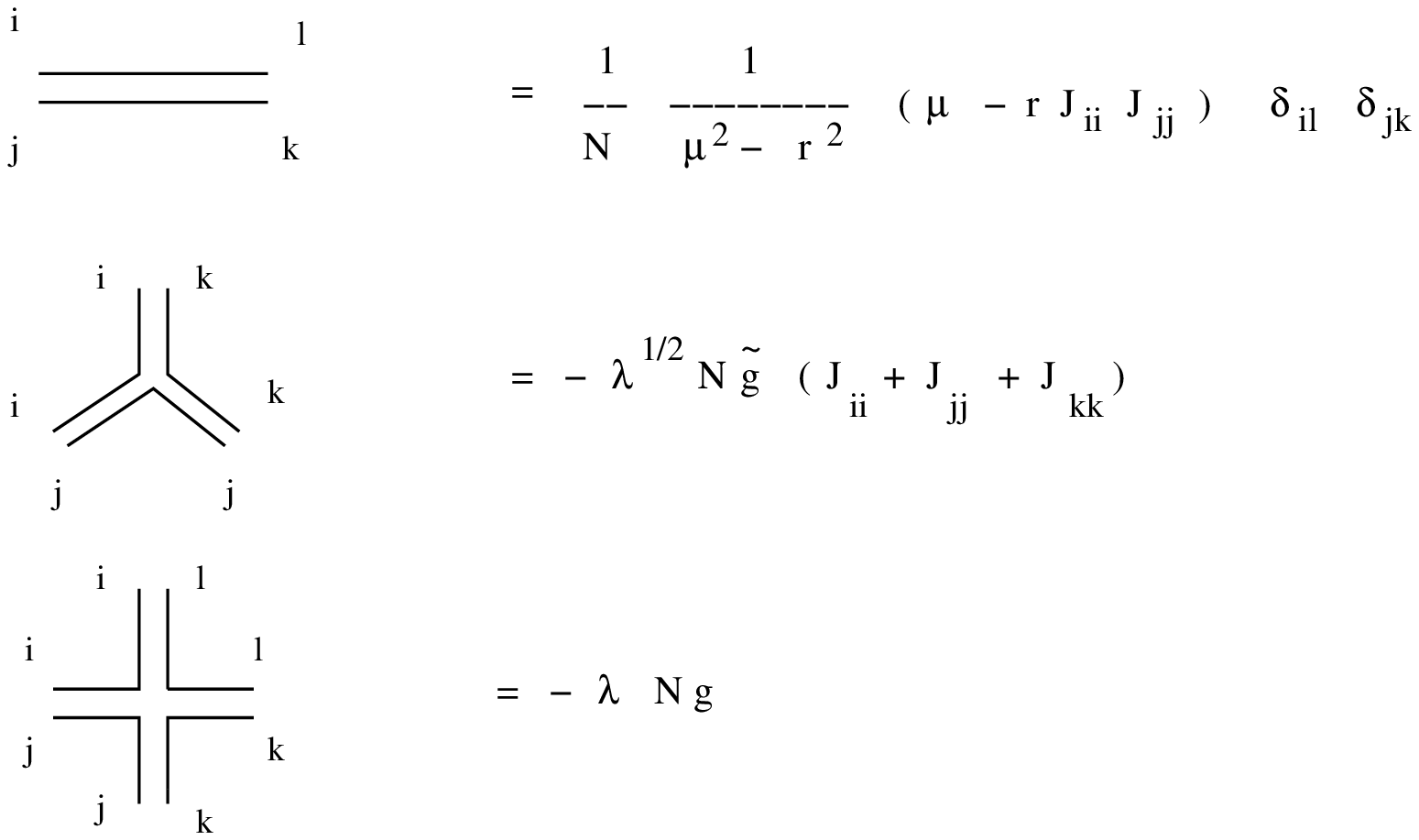,height=8cm}
    \caption{The Feynman rules for the $\lambda$-expansion 
of (\ref{ZJphi34B}).}
    \label{fig:feynman}
}

\subsection{Full propagator, tadpole and 2PI vacuum diagrams}
As the fundamental quantities in the present calculation, 
we consider the full propagator (the connected two-point function) 
and the tadpole (the one-point function),
which can be written respectively as
\beqa
\langle \phi_{ij}\phi_{kl}\rangle_{\rm C} & \equiv & 
\langle \phi_{ij}\phi_{kl}\rangle - 
\langle \phi_{ij} \rangle\langle \phi_{kl} \rangle
= \frac1N (c + d J_{ii}J_{jj})\delta_{il}\delta_{jk} \ , \n 
\langle \phi_{ij}\rangle & = & \sqrt{\lambda}\, T J_{ii}\delta_{ij} 
\eeqa
due to the SU($N/2$)$\times$SU($N/2$) symmetry.
At the leading order, $c$, $d$ and $T$ are given by
\beq
c = \frac{\mu}{\mu^2 - r^2} + O(\lambda)\ , \qquad
d = \frac{-r}{\mu^2 - r^2} + O(\lambda) \ , \qquad
T = -\tilde{g}\frac{1}{\mu + r}\frac{2\mu-r}{\mu^2-r^2} + O(\lambda) \ . 
\label{leading_cdT}
\eeq

In what follows we will use the Schwinger-Dyson equations
for the full propagator and the tadpole
to reduce the calculation of the free energy 
to that of two-particle-irreducible (2PI) planar vacuum diagrams.
By `two-particle-irreducible' we mean
that the diagram cannot be separated into two parts 
by removing two propagators. 
For later convenience we use the full propagator for internal 
lines, and denote the sum of those diagrams as $N^2 G(c,d)$.
For example $N^2 G(c,d)$ consists of 5 diagrams shown in 
Fig.\ \ref{fig:2PI} up to the 2nd order,
and the results of the first two diagrams are given by 
$\frac12\lambda\tilde{g}^2N^2(c^3 + d^3 + 2c^2d + 2cd^2)$ and 
$-\frac12\lambda gN^2c^2$ respectively. 
The important point is that there are much less
two-particle-irreducible (2PI) vacuum diagrams 
than general vacuum diagrams that need to be considered
in order to obtain the free energy ${\cal F}(\mu , r)$ directly.

\FIGURE{\epsfig{file=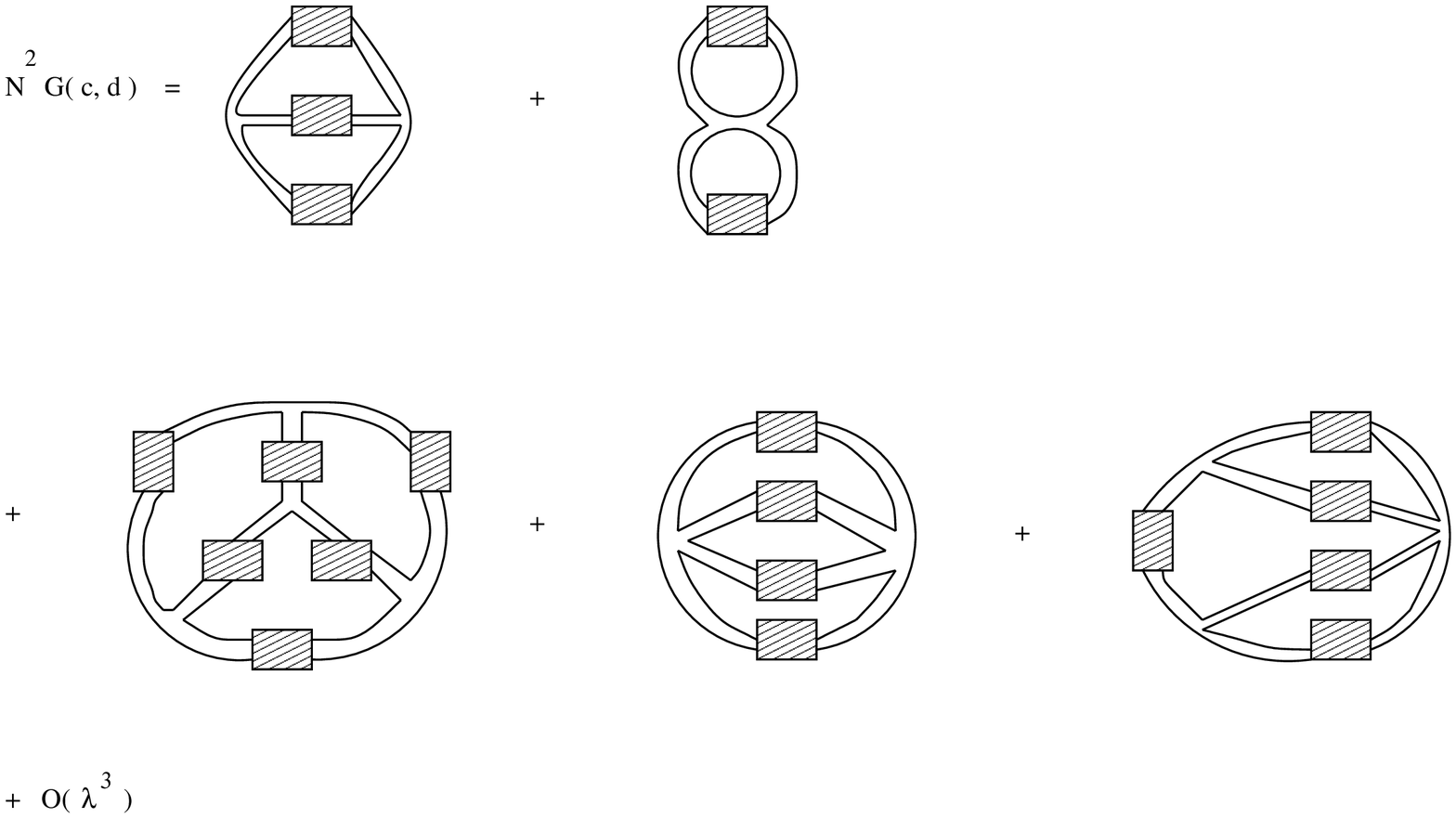,height=8cm}
    \caption{The 2PI planar vacuum diagrams
up to $O(\lambda^2)$.
The internal line with the rectangular blob stands for 
the full propagator. 
The two diagrams in the first line are of $O(\lambda)$, while the three 
diagrams in the second line are of $O(\lambda^2)$.  }
    \label{fig:2PI}
}

\subsection{Derivation of the Schwinger-Dyson equations}
Here we will derive the Schwinger-Dyson equations,
which allow us
to obtain the full propagator and the tadpole order by order in 
$\lambda$.

\FIGURE{\epsfig{file=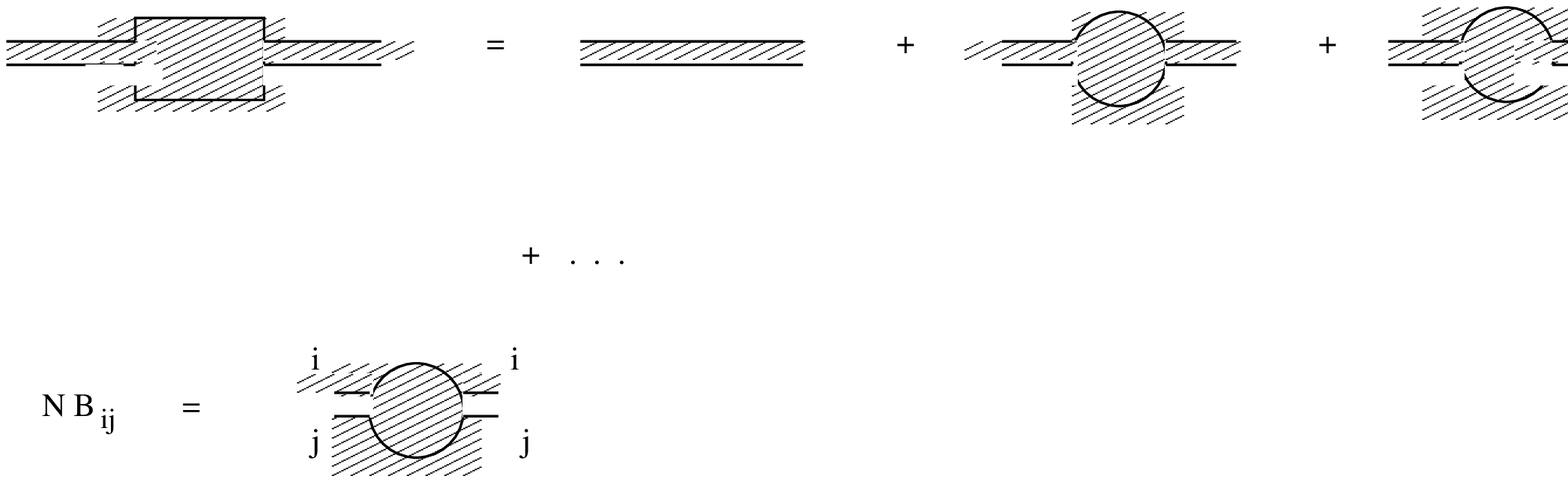,width=14cm}
    \caption{The full propagator can be expressed as a geometric series.
The round blob represents
the radiative correction to the quadratic term,
which is denoted by $NB_{ij}$ as indicated in the second line.}
    \label{fig:radiative}
}
\FIGURE{\epsfig{file=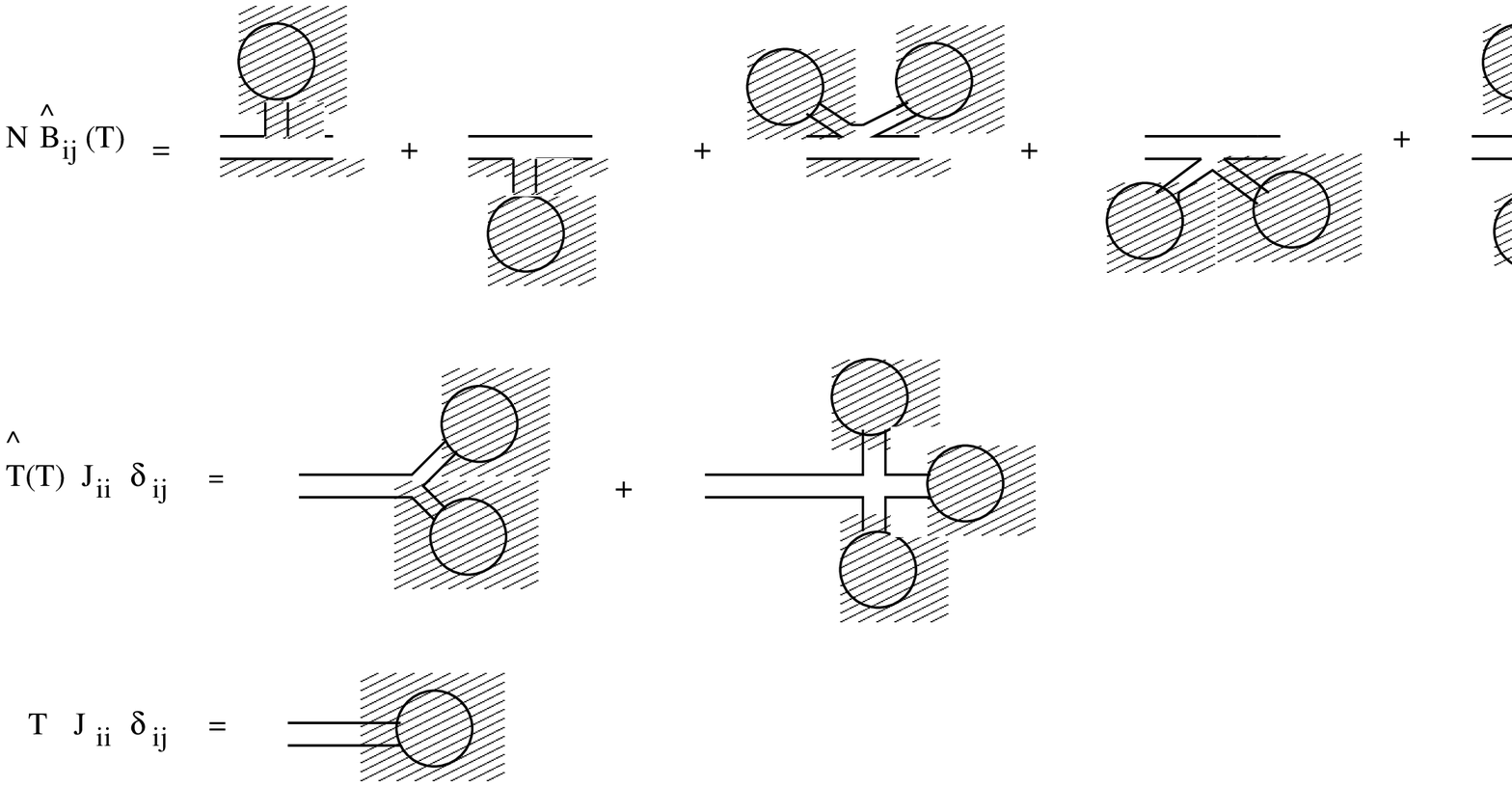,width=14cm}
    \caption{The first two lines show the diagrams which contribute
to the parts of $B_{ij}$ and $T$ which depend only on $T$ 
but not on $c$ or $d$.
The round blob attached at the ends of the double-line 
represents the tadpole $TJ_{ii}\delta_{ij}$ as stated in the last line.}
    \label{fig:radiativeBT}
}
\FIGURE{\epsfig{file=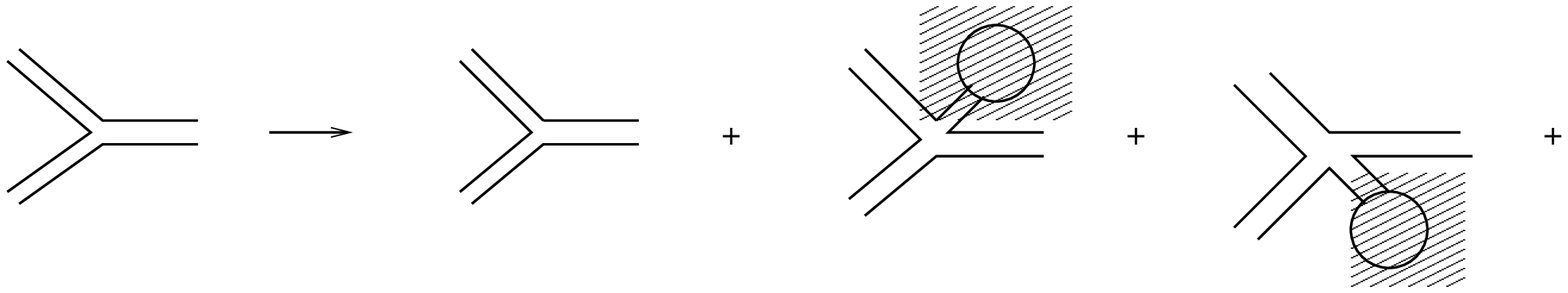,width=14cm}
    \caption{The full result for $\hat{B}_{ij}(c,d,T)$ can be
obtained from $\hat{B}_{ij}(c,d,T=0)$ by dressing
the 3-point vertices with the tadpole. }
    \label{fig:replacement}
}
\FIGURE{\epsfig{file=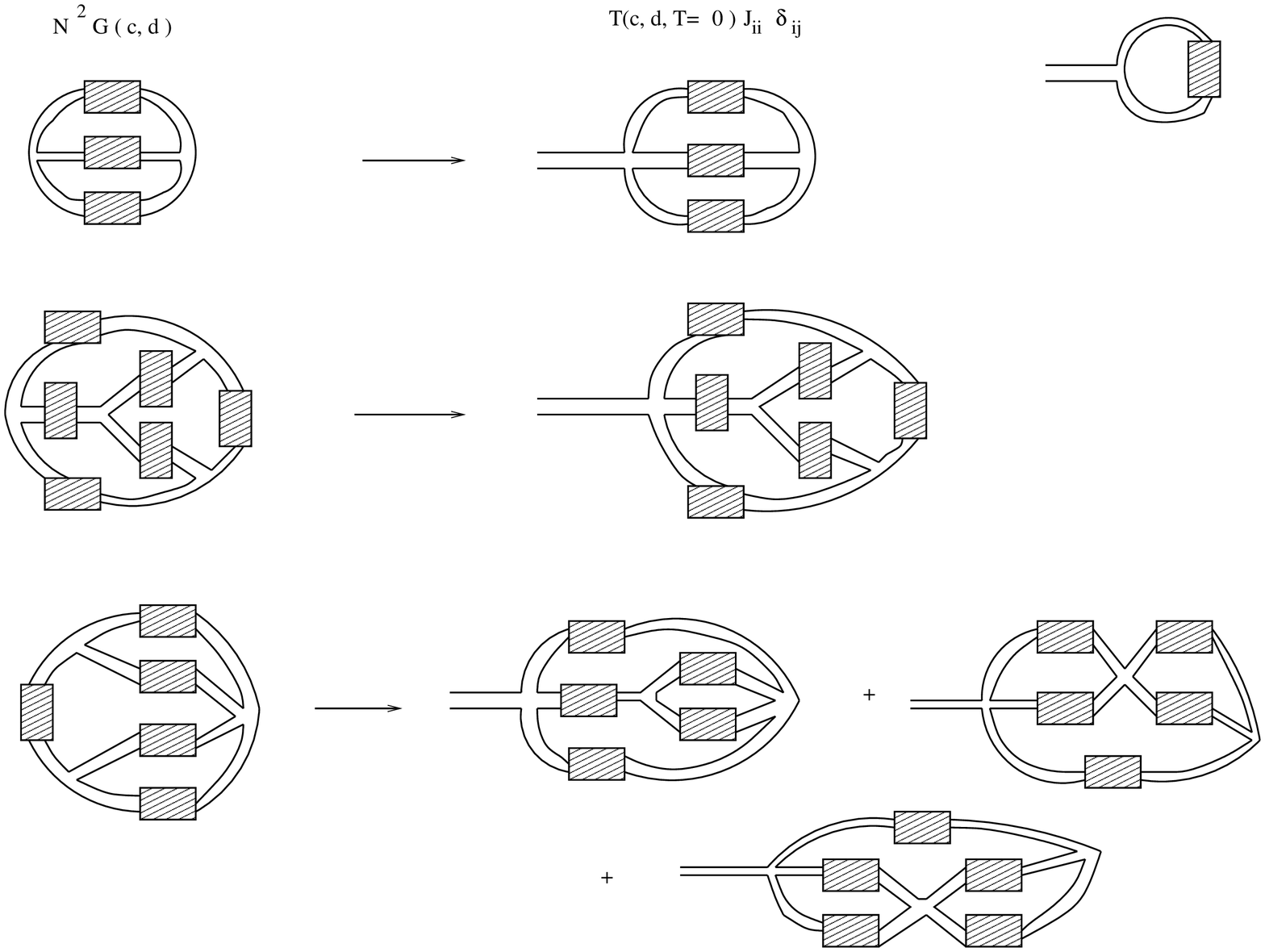,width=14cm}
    \caption{Except for the lowest order diagram shown
in the upper right corner, 
all the diagrams contributing to $\hat{T}(c,d,T=0)$ are obtained 
by attaching an external line to a 3-point vertex in the
diagrams contributing to $N^2G(c,d)$. 
The diagrams at the first few orders are shown for illustration. }
    \label{fig:2PI_to_tad}
}

Let us note first that
the full propagator can be expressed as a geometric series
as depicted in Fig.\ \ref{fig:radiative}.
The round blob, which we denote as $N B_{ij}$,
stands for the radiative correction to the quadratic term.
One can solve this relation for $B_{ij}$ as 
\beq
B_{ij} = \mu - \frac{c}{c^2-d^2} + 
\left( r + \frac{d}{c^2 - d^2}\right)J_{ii}J_{jj} \ . 
\label{radiative1}
\eeq 
It is convenient to decompose $B_{ij}$ as
\beq
B_{ij} = \hat{B}_{ij}(T) + \hat{B}_{ij}(c,d,T) \ , 
\label{decomp_B}
\eeq
where $\hat{B}_{ij}(T)$ contains only tadpoles and hence do not depend
on $c$ or $d$. One can easily see that $\hat{B}_{ij}(T)$ consists of 
5 diagrams shown in Fig.\ \ref{fig:radiativeBT}, which yield
\beq
\label{BijT}
\hat{B}_{ij}(T) = (-2\lambda \tilde{g}T -\lambda^2 gT^2)(2 + J_{ii}J_{jj})
\ . 
\eeq
The second term in (\ref{decomp_B}) can be obtained from 
the sum of 2PI diagrams $G(c,d)$ as follows.
Note first that $\hat{B}_{ij}(c,d,T=0)$
can be obtained from 
$G(c,d)$ by removing one full propagator, i.e.,
\beq
\hat{B}_{ij}(c,d,T=0) = 2\frac{\partial}{\partial c}G(c,d) 
+ 2 \frac{\partial}{\partial d}G(c,d) J_{ii}J_{jj} \ . 
\label{hatB_Tzero}
\eeq
Then one can obtain $\hat{B}_{ij}(c,d,T)$ for $T\neq 0$
by the shifting the three-point vertex as (See Fig.\ \ref{fig:replacement})
\beq
\hat{B}_{ij}(c,d,T) = \left. \hat{B}_{ij}(c,d,T=0)
\right|_{\tilde{g}\rightarrow \tilde{g} + \lambda gT} \ . 
\label{BijcdT}
\eeq 
Summation of (\ref{BijT}) and (\ref{BijcdT}) yields
\beq
B_{ij} =   \left[2\frac{\partial}{\partial c}G(c,d) 
+ 2 \frac{\partial}{\partial d}G(c,d) J_{ii}J_{jj}
\right]_{\tilde{g}\rightarrow \tilde{g} + \lambda gT} 
- (2\lambda \tilde{g}T +\lambda^2 gT^2)(2 + J_{ii}J_{jj}) \ .
\label{radiative2}
\eeq
From (\ref{radiative1}) and (\ref{radiative2}) we obtain
the Schwinger-Dyson equations for the full propagator
\beqa
 & & \mu - \frac{c}{c^2-d^2} =  \left[ 2\frac{\partial}{\partial c}G(c,d)
\right]_{\tilde{g}\rightarrow \tilde{g} + \lambda gT}  
 -2 (2\lambda \tilde{g}T +\lambda^2 gT^2) \ , \n 
 & & r + \frac{d}{c^2-d^2} =  \left[ 2\frac{\partial}{\partial d}G(c,d)
\right]_{\tilde{g}\rightarrow \tilde{g} + \lambda gT}  
 -(2\lambda \tilde{g}T +\lambda^2 gT^2) \ . 
\label{SD1}
\eeqa

The Schwinger-Dyson equation for the tadpole
can be obtained in a similar way. 
Let us first decompose the tadpole $T$ as 
\beq
T = \hat{T}(T) + \hat{T}(c,d,T) \ . 
\label{SD_tad1}
\eeq
The first term, which includes only tadpoles,
consists of two diagrams shown in Fig.\ \ref{fig:radiativeBT}, 
which yield
\beq
\hat{T}(T) = \frac{1}{\mu + r}(-3\lambda\tilde{g}T^2 -\lambda^2 gT^3) \ . 
\label{SD_tad2}
\eeq
Next we consider the diagrams that contribute to $\hat{T}(c,d,T=0)$.
Note that every diagram other than the one at the lowest order
can be obtained from 2PI vacuum diagrams
by attaching an external line to a 3-point vertex
as illustrated in Fig.\ \ref{fig:2PI_to_tad}. 
Thus we have
\beq
\hat{T}(c,d,T=0) = -\tilde{g}\frac{1}{\mu + r}(2c+d) + 
g\frac{1}{\mu+r}\frac{\partial}{\partial \tilde{g}}G(c,d) \ ,  
\eeq
where the first term represents the contribution from the lowest order 
diagram. 
Then $\hat{T}(c,d,T)$ with $T\neq 0$
can be obtained from $\hat{T}(c,d,T=0)$
by the same replacement as in Fig.\ \ref{fig:replacement},
which leads us to
\beq
\hat{T}(c,d,T) = \left.\frac{1}{\mu + r}\left[-\tilde{g}(2c+d)
+g \frac{\partial}{\partial \tilde{g}}G(c,d)\right]
\right|_{\tilde{g}\rightarrow \tilde{g} + \lambda gT}. 
\label{TcdT}
\eeq
From (\ref{SD_tad1}), (\ref{SD_tad2}) and  (\ref{TcdT}), we 
obtain the Schwinger-Dyson equation for the tadpole
\beq
T  =  \left.\frac{1}{\mu + r}\left[-\tilde{g}(2c+d)
+g \frac{\partial}{\partial \tilde{g}}G(c,d)\right]
\right|_{\tilde{g}\rightarrow \tilde{g} + \lambda gT} 
-\frac{1}{\mu + r}(3\lambda\tilde{g}T^2 +\lambda^2 gT^3)  \ . 
\label{SD2}
\eeq

Since Eqs.\ (\ref{SD1}) and (\ref{SD2}) are closed with respect to 
$c$, $d$, $T$, we can solve them order by order
once the sum of 2PI diagrams $G(c,d)$ is obtained.

\subsection{Explicit form of the free energy}
From the full propagator and the tadpole,
one can readily obtain the expectation value of 
the quadratic terms in (\ref{ZJphi34B}) as
\beq
{\cal M} \equiv  \left\langle\frac1N \tr \phi^2 \right\rangle 
= c + \lambda T^2, \qquad
{\cal M}_J \equiv \left\langle\frac1N \tr (J\phi J\phi) \right\rangle 
= d + \lambda T^2. 
\eeq
Since the free energy ${\cal F}(\mu , r)$ satisfies
the differential equations
\beq
4\frac{\partial}{\partial \mu_+}{\cal F}(\mu, r) = {\cal M} + {\cal M}_J
~~~,~~~ 
\qquad
4\frac{\partial}{\partial \mu_-}{\cal F}(\mu, r) = {\cal M} - {\cal M}_J
\ ,  
\label{fe_bibun}
\eeq
where $\mu_{\pm} = \mu \pm r$,
we can obtain the $\lambda$-expansion of 
the free energy by term-by-term integration. 
The integration constant in ${\cal F}(\mu, r)$, which does not
depend on $\mu_{\pm}$, appears only at the order of $O(\lambda^0)$, 
and it can be calculated directly as $\frac12(\ln N -\ln 2)$. 

The number of 2PI planar vacuum diagrams up to the 3rd order is 13.
By evaluating them explicitly and by following the above procedure
we obtain the explicit form of the free energy ${\cal F}(\mu, r)$ 
up to the 3rd order as follows.
\beqa
{\cal F}(\mu, r) & = & \sum_{n = 0}^{\infty}\lambda^n{\cal F}_n(\mu, r), \\
{\cal F}_0(\mu, r) & = &  \frac14 
\Bigl[ 2\ln N + \ln(\mu_+\mu_-) -2\ln 2 \Bigr], \n
{\cal F}_1(\mu, r) & = & \frac{\tilde{g}^2}{4}
\left(-\frac{6}{\mu_+^3}-\frac{3}{\mu_+^2\mu_-}-\frac{1}{\mu_+\mu_-^2}\right)
+\frac{g}{4}\left(\frac{1}{2\mu_+^2}+\frac{1}{\mu_+\mu_-}+\frac{1}{2\mu_-^2}
\right), \n
{\cal F}_2(\mu, r) & = & \frac{\tilde{g}^4}{4}\left(-\frac{108}{\mu_+^6}
-\frac{54}{\mu_+^5\mu_-} - \frac{81}{4\mu_+^4\mu_-^2}
-\frac{17}{2\mu_+^3\mu_-^3} - \frac{9}{4\mu_+^2\mu_-^4}\right)\n
 & & +\frac{g\tilde{g}^2}{4}\left(\frac{27}{\mu_+^5} 
+ \frac{99}{4\mu_+^4\mu_-}+\frac{47}{4\mu_+^3\mu_-^2} 
+ \frac{25}{4\mu_+^2\mu_-^3} + \frac{9}{4\mu_+\mu_-^4}\right) \n
 & & +\frac{g^2}{4}\left(-\frac{9}{16\mu_+^4} -\frac{1}{\mu_+^3\mu_-}
-\frac{11}{8\mu_+^2\mu_-^2}-\frac{1}{\mu_+\mu_-^3}-\frac{9}{16\mu_-^4}
\right), \n 
{\cal F}_3(\mu, r) & = & \frac{\tilde{g}^6}{4}\left(-\frac{3402}{\mu_+^9}
-\frac{1944}{\mu_+^8\mu_-}-\frac{810}{\mu_+^7\mu_-^2}
-\frac{351}{\mu_+^6\mu_-^3}-\frac{135}{\mu_+^5\mu_-^4}
-\frac{45}{\mu_+^4\mu_-^5}-\frac{9}{\mu_+^3\mu_-^6}\right) \n
 & & +\frac{g\tilde{g}^4}{4}\left(\frac{1377}{\mu_+^8}
+\frac{1188}{\mu_+^7\mu_-}+\frac{1215}{2\mu_+^6\mu_-^2}
+\frac{306}{\mu_+^5\mu_-^3}+\frac{135}{\mu_+^4\mu_-^4}
+\frac{54}{\mu_+^3\mu_-^5} + \frac{27}{2\mu_+^2\mu_-^6}\right) \n
 & & + \frac{g^2\tilde{g}^2}{4}\left(-\frac{243}{2\mu_+^7}
-\frac{621}{4\mu_+^6\mu_-}-\frac{477}{4\mu_+^5\mu_-^2}
-\frac{153}{2\mu_+^4\mu_-^3}-\frac{81}{2\mu_+^3\mu_-^4}
-\frac{81}{4\mu_+^2\mu_-^5}-\frac{27}{4\mu_+\mu_-^6}\right) \n
 & & +\frac{g^3}{4}\left(\frac{9}{8\mu_+^6}+\frac{9}{4\mu_+^5\mu_-}
+\frac{27}{8\mu_+^4\mu_-^2} + \frac{9}{2\mu_+^3\mu_-^3}
+ \frac{27}{8\mu_+^2\mu_-^4} + \frac{9}{4\mu_+\mu_-^5} + \frac{9}{8\mu_-^6}
\right) \ .  \nonumber 
\eeqa




\begin{thebibliography}{999}




\bibitem{conv} 
W.~E.~Caswell,
{\em Accurate Energy Levels For The Anharmonic Oscillator 
And A Summable Series For The Double Well Potential In  Perturbation Theory},
Annals Phys.\  {\bf 123} (1979) 153;
I.~G.~Halliday and P.~Suranyi,
{\em The Anharmonic Oscillator: A New Approach},
Phys.\ Rev.\ D {\bf 21} (1980) 1529;
J.~Killingbeck, {\em Renormalised Perturbation Series}, 
J.\ Phys.\ A {\bf 14} (1981) 1005.


\bibitem{Stevenson:1981vj}
P.~M.~Stevenson,
{\em Optimized Perturbation Theory},
Phys.\ Rev.\ D {\bf 23} (1981) 2916 .



\bibitem{exact_conv}
P.~M.~Stevenson,
{\em Optimization And The Ultimate Convergence Of QCD Perturbation Theory},
Nucl.\ Phys.\ B {\bf 231} (1984) 65; 
I.~R.~Buckley, A.~Duncan and H.~F.~Jones,
{\em Proof Of The Convergence Of The Linear Delta Expansion}, 
Phys.\ Rev.\ D {\bf 47} (1993) 2554; 
A.~Duncan and H.~F.~Jones,
{\em Convergence proof for optimized Delta expansion: The Anharmonic oscillator}, 
{\em ibid.} (1993) 2560;
C.~M.~Bender, A.~Duncan and H.~F.~Jones,
{\em Convergence of the optimized delta expansion for 
the connected vacuum amplitude: Zero dimensions},
Phys.\ Rev.\ D {\bf 49} (1994) 4219
[{\tt hep-th/9310031}]; 
R.~Guida, K.~Konishi and H.~Suzuki,
{\em Convergence of scaled delta expansion: Anharmonic oscillator},
Annals Phys.\  {\bf 241} (1995) 152
[{\tt hep-th/9407027}]; 
{\em Improved convergence proof of the delta expansion and order dependent mappings}, 
{\em ibid.}  {\bf 249} (1996) 109
[{\tt hep-th/9505084}].


\bibitem{GEM_field}
A.~Okopinska,
{\em Nonstandard Expansion Techniques For The Effective Potential 
In $\lambda\Phi^4$ Quantum Field Theory}, 
Phys.\ Rev.\ D {\bf 35} (1987) 1835;
{\em Optimized Expansion In Quantum Field Theory Of Massive Fermions With 
$(\bar{\psi}\psi)^2$ Interaction}, 
Phys.\ Rev.\ D {\bf 38} (1988) 2507; 
I.~Stancu and P.~M.~Stevenson,
{\em Second Order Corrections to the Gaussian 
Effective Potential of $\lambda \phi^4$ Theory},
Phys.\ Rev.\ D {\bf 42} (1990) 2710;
E.~Braaten and E.~Radescu,
{\em Convergence of the linear delta expansion 
in the critical O(N) field theory}, 
[{\tt hep-ph/0206108}].


\bibitem{Dhar:sh}
A.~Dhar,
{\em Renormalization Scheme - Invariant Perturbation Theory}, 
Phys.\ Lett.\ B {\bf 128} (1983) 407.

\bibitem{Kawamoto:2003kn}
S.~Kawamoto and T.~Matsuo,
{\em Improved renormalization group analysis for Yang-Mills theory},
[{\tt hep-th/0307171}].


\bibitem{Kabat:2000hp}
D.~Kabat and G.~Lifschytz,
{\em Approximations for strongly-coupled supersymmetric quantum
mechanics},
Nucl.\ Phys.\ {\bf B 571} (2000) 419
[{\tt hep-th/9910001}].

\bibitem{Banks:1996vh}
T.~Banks, W.~Fischler, S.~H.~Shenker and L.~Susskind,
{\em M theory as a matrix model: A conjecture}, 
Phys.\ Rev.\ D {\bf 55} (1997) 5112
[{\tt hep-th/9610043}].


\bibitem{blackholes}
D.~Kabat, G.~Lifschytz and D.A.~Lowe,
{\em Black hole thermodynamics from calculations 
in strongly-coupled gauge theory},
Phys.\ Rev.\ Lett.\  {\bf 86} (2001) 1426
[{\tt hep-th/0007051}];
%
{\em Black hole entropy from non-perturbative gauge theory},
Phys.\ Rev.\ D {\bf 64} (2001) 124015
[{\tt hep-th/0105171}];
%
N.~Iizuka, D.~Kabat, G.~Lifschytz and D.~A.~Lowe,
{\em Probing black holes in non-perturbative gauge theory},
Phys.\ Rev.\ D {\bf 65} (2002) 024012
[{\tt hep-th/0108006}].


\bibitem{EL}
M.~Engelhardt and S.~Levit,
{\em Variational master field for large-N interacting matrix models: 
Free random variables on trial}, 
Nucl.\ Phys.\ B {\bf 488} (1997) 735
[{\tt hep-th/9609216}].





\bibitem{Nishimura:2001sx}
J.~Nishimura and F.~Sugino,
{\em Dynamical generation of four-dimensional space-time 
in the IIB matrix model},
JHEP {\bf 0205} (2002) 001 
[{\tt hep-th/0111102}].



\bibitem{IKKT}N.\ Ishibashi, H.\ Kawai, Y.\ Kitazawa and A.\ Tsuchiya,
{\em A Large-$N$ Reduced Model as Superstring},
Nucl.\ Phys. {\bf B~498} (1997) 467 [{\tt hep-th/9612115}].

\bibitem{Austing:2001bd}
P.~Austing and J.F.~Wheater,
{\em The convergence of Yang-Mills integrals},
JHEP {\bf 0102} (2001) 028 
[{\tt hep-th/0101071}];
{\em Convergent Yang-Mills matrix theories}, 
JHEP {\bf 0104} (2001) 019
[{\tt hep-th/0103159}].


\bibitem{HNT} 
T.\ Hotta, J.\ Nishimura and A.\ Tsuchiya, 
{\em Dynamical Aspects of Large $N$ Reduced Models},
Nucl.\ Phys. {\bf B~545} (1999) 543
[{\tt hep-th/9811220}].
%

\bibitem{monte}
W.~Krauth, H.~Nicolai and M.~Staudacher,
{\em Monte Carlo approach to M-theory},
Phys.\ Lett.\ B {\bf 431} (1998) 31
[{\tt hep-th/9803117}];
W.~Krauth and M.~Staudacher,
{\em Finite Yang-Mills integrals},
Phys.\ Lett.\ {\bf B 435} (1998) 350 
[{\tt hep-th/9804199}];
J.\ Ambj\o rn, K.N.\ Anagnostopoulos, W.\ Bietenholz,
T.\ Hotta and J.\ Nishimura,
{\em Large $N$ Dynamics of Dimensionally Reduced 4D SU($N$) Super Yang-Mills 
Theory}, 
JHEP {\bf 0007} (2000) 013 [{\tt hep-th/0003208}];
J.\ Ambj\o rn, K.N.\ Anagnostopoulos, W.\ Bietenholz,
T.\ Hotta and J.\ Nishimura,
{\em Monte Carlo Studies of the IIB Matrix Model at Large $N$},
JHEP {\bf 0007} (2000) 011 [{\tt hep-th/0005147}];
P. Bialas, Z. Burda, B. Petersson and J. Tabaczek,
{\em Large $N$ Limit of the IKKT Model},
Nucl. Phys. {\bf B 592} (2001) 391 [{\tt hep-lat/0007013}];
Z.~Burda, B.~Petersson and J.~Tabaczek,
{\em Geometry of Reduced Supersymmetric 4D Yang-Mills Integrals},
Nucl.\ Phys.\ {\bf B 602} (2001) 399 [{\tt hep-lat/0012001}];
J.~Ambj\o rn, K.N.~Anagnostopoulos, 
W.~Bietenholz, F.~Hofheinz and J.~Nishimura,
{\em On the Spontaneous Breakdown of Lorentz Symmetry 
in Matrix Models of Superstrings}, 
Phys.\ Rev.\ D {\bf 65} (2002) 086001 [{\tt hep-th/0104260}];
K.N. Anagnostopoulos, W. Bietenholz and J. Nishimura,
{\em The Area Law in Matrix Models for Large $N$ QCD Strings},
Int.\ J.\ Mod.\ Phys.\ C {\bf 13} (2002) 555
[{\tt hep-lat/0112035}].



\bibitem{Gauss_simpleIIB}
S.~Oda and F.~Sugino,
{\em Gaussian and mean field approximations 
for reduced Yang-Mills integrals},
JHEP {\bf 0103} (2001) 026
[{\tt hep-th/0011175}];
F.~Sugino,
{\em Gaussian and mean field approximations for reduced 4D supersymmetric 
Yang-Mills integral},
JHEP {\bf 0107} (2001) 014 
[{\tt hep-th/0105284}].



\bibitem{Aoki:1998vn}
H.~Aoki, S.~Iso, H.~Kawai, Y.~Kitazawa and T.~Tada,
{\em Space-time structures from IIB matrix model}, 
Prog.\ Theor.\ Phys.\  {\bf 99} (1998) 713
[{\tt hep-th/9802085}].




\bibitem{NV}
J.\ Nishimura and G.\ Vernizzi,
{\em Spontaneous Breakdown of Lorentz Invariance in IIB Matrix Model},
JHEP {\bf 0004} (2000) 015 
[{\tt hep-th/0003223}];
{\em Brane World Generated Dynamically from String Type IIB Matrices},
Phys. Rev. Lett. {\bf 85} (2000) 4664 
[{\tt hep-th/0007022}];
J.~Nishimura,
{\em Exactly solvable matrix models for the dynamical generation of  
space-time in superstring theory},
Phys.\ Rev.\ D {\bf 65} (2002) 105012
[{\tt hep-th/0108070}].



\bibitem{sign}
K.~N.~Anagnostopoulos and J.~Nishimura,
{\em New approach to the complex-action problem 
and its application to a nonperturbative study of superstring theory},
Phys.\ Rev.\ D {\bf 66} (2002) 106008
[{\tt hep-th/0108041}].


\bibitem{Vernizzi:2002mu}
G.~Vernizzi and J.~F.~Wheater,
{\em Rotational symmetry breaking in multi-matrix models},
Phys.\ Rev.\ D {\bf 66} (2002) 085024
[{\tt hep-th/0206226}].




\bibitem{Imai:2003jb}
T.~Imai, Y.~Kitazawa, Y.~Takayama and D.~Tomino,
{\em Effective actions of matrix models on homogeneous spaces},
[{\tt hep-th/0307007}].



\bibitem{KKKMS}
H.~Kawai, S.~Kawamoto, T.~Kuroki, T.~Matsuo and S.~Shinohara,
{\em Mean field approximation of IIB matrix model and 
emergence of four dimensional space-time},
Nucl.\ Phys.\ B {\bf 647} (2002) 153
[{\tt hep-th/0204240}].


\bibitem{Kawai:2002ub}
H.~Kawai, S.~Kawamoto, T.~Kuroki and S.~Shinohara,
{\em Improved perturbation theory and four-dimensional 
space-time in IIB  matrix model},
Prog.\ Theor.\ Phys.\  {\bf 109} (2003) 115
[{\tt hep-th/0211272}].


\bibitem{Nishimura:2002va}
J.~Nishimura, T.~Okubo and F.~Sugino,
{\em Convergence of the Gaussian expansion method
in dimensionally reduced Yang-Mills integrals},
JHEP {\bf 0210} (2002) 043
[{\tt hep-th/0205253}].



\bibitem{Brezin:1977sv}
E.~Brezin, C.~Itzykson, G.~Parisi and J.~B.~Zuber,
{\em Planar Diagrams},
Commun.\ Math.\ Phys.\  {\bf 59} (1978) 35.




\bibitem{Cicuta:1986pu}
G.~M.~Cicuta, L.~Molinari and E.~Montaldi,
{\em Large N Phase Transitions In Low Dimensions},
Mod.\ Phys.\ Lett.\ A {\bf 1} (1986) 125.





\end{thebibliography}
\end{document}